\documentclass[11pt,showpacs,superscriptaddress,amsmath,amssymb,nofootinbib]{revtex4}
\usepackage{graphicx}

\begin{document}

\bigskip

\title{Properties and searches of the exotic neutral Higgs bosons in the Georgi-Machacek model}

\author{Cheng-Wei Chiang}
\email[e-mail: ]{chengwei@ncu.edu.tw}
\affiliation{Center for Mathematics and Theoretical Physics and Department of
  Physics, National Central University, Chungli, Taiwan 32001, R.O.C.}
\affiliation{Institute of Physics, Academia Sinica, Taipei, Taiwan 11529, R.O.C.}
\affiliation{Physics Division, National Center for Theoretical Sciences, Hsinchu, Taiwan 30013, R.O.C.}
\affiliation{Kobayashi-Maskawa Institute for the Origin of Particles and the Universe, Nagoya University, Nagoya 464-8602, Japan}

\author{Koji Tsumura}
\email[e-mail: ]{ko2@gauge.scphys.kyoto-u.ac.jp}
\affiliation{Department of Physics, Kyoto University, Kyoto 606-8502, Japan}

\date{\today}

\begin{abstract}

The Georgi-Machacek model predicts the existence of four neutral Higgs bosons, one of which can be identified as the 125-GeV Higgs boson.  The latest Higgs data favor the parameter space of small mixing angle $\alpha$ between the two custodial singlets of the model. The other two neutral Higgs bosons belong respectively to the custodial triplet and quintet. We study the general decay and production properties of these particles in the small-$\alpha$ scenario.  Constraints on the $SU(2)_L$ triplet vacuum expectation value are obtained as a function of the exotic Higgs boson masses using latest ATLAS data of various search channels for additional neutral Higgs bosons.

\end{abstract}


\preprint{KUNS-2537}



\maketitle

\newpage

\section{Introduction \label{sec:intro}}

For almost half a century since its introduction to particle physics, the standard model (SM) has survived many precision tests.  Over the years, we have attained to knowing sufficient details of its structure and parameters from experimental data.  Despite its success in explaining countless empirical observations, the SM cannot be a UV complete theory.  From a theoretical point of view, the model is unsatisfactory for various reasons.  For example, the Higgs boson mass is found to be at the electroweak scale and its radiative correction poses a fine-tuning issue.  There are also unanswered questions such as why fermions of different flavors have such a mass hierarchy and what is the origin of CP violation in the quark sector.  Experimentally, we have found the phenomena of neutrino oscillations and dark matter/energy in the Universe that the SM falls short of.

The discovery of the SM-like Higgs boson at the mass of about 125 GeV not only completes the particle spectrum in the SM, but also stimulates studies about detailed properties of the particle and pursuits of an extended Higgs family.  Among many new physics models with a larger Higgs sector, the Georgi-Machacek (GM) model~\cite{Georgi:1985nv,Chanowitz:1985ug} has received much attention in view of its intriguing features.  The Higgs sector of GM model houses a complex triplet of hypercharge $Y = 1$ and a real triplet of $Y = 0$ under the SM $SU(2)_L \times U(1)_Y$ gauge symmetry, in addition to a SM-like doublet.  
With vacuum alignment between the complex and real triplets, the model preserves the custodial symmetry at tree level, granting the possibility of a triplet vacuum expectation value (VEV) as large as up to a few tens of GeV.  Naturalness associated with divergent one-loop corrections to the electroweak $\rho$ parameter and certain mixings among the Higgs bosons have been studied and found to be similar to the SM Higgs mass~\cite{Gunion:1990dt}.  The model predicts the existence of many Higgs bosons, forming two singlets, one triplet, and one quintet under the custodial symmetry~\cite{Gunion:1989ci}.  There have been extensive phenomenological studies of the exotic Higgs bosons in the literature~\cite{Haber:1999zh,Godfrey:2010qb,Chiang:2012dk,Chiang:2012cn,Englert:2013zpa,Englert:2013wga,Chiang:2013rua,Hartling:2014zca,Chiang:2014bia,Hartling:2014aga}, including their effects in enhancing the strength of phase transition in electroweak baryogenesis~\cite{Chiang:2014hia}.  Due to mixing between the doublet and the triplet fields, the coupling between the SM-like Higgs boson and the weak gauge bosons can be greater than the SM value~\cite{Logan:2010en,Falkowski:2012vh,Chang:2012gn,Chiang:2013rua}, which is impossible for models with only extra $SU(2)_L$ singlet and/or doublet fields.~\footnote{
Mixing with an $SU(2)_L$ septet scalar field with $Y=2$ is another possibility to obtain $\kappa_V^{} > 1$.  Such a model also predicts $\rho = 1$ without the need of vacuum alignment\cite{Hisano:2013sn}.}
Besides, the model has a flavor-universal correction to the Yukawa couplings of the SM-like Higgs boson.  Although some of the above-mentioned features are also shared by the septet extension model~\cite{Hisano:2013sn}, the latter suffers from an accidental $U(1)$ symmetry that could lead to an undesirable Nambu-Goldstone (NG) boson if it is broken spontaneously.  There are also supersymmetric~\cite{Cort:2013foa,Garcia-Pepin:2014yfa} and little Higgs extensions~\cite{Chang:2003un,Chang:2003zn} of the GM model.

In this paper, we concentrate on the productions and decays of the neutral Higgs bosons in the GM model at the CERN Large Hadron Collider (LHC).  After the discovery of the 125-GeV Higgs boson, it is of great interest to know whether there exists another neutral Higgs boson of higher or even lower mass.  Therefore, continued efforts are made to search for such Higgs-like resonances within the full mass range explorable at the LHC, utilizing various major production and decay channels.  Such searches offer the exclusion limit in each channel as a function of the Higgs mass.  In the GM model, there are totally four neutral Higgs bosons.  We use $h$, $H_1^0$ to denote respectively the SM-like Higgs boson discovered at the LHC and the other Higgs singlet.  Both of them are mixtures of the CP-even custodial singlets.  The other two are the CP-odd $H_3^0$ belonging to the triplet and the CP-even $H_5^0$ in the quintet.  Since $h$ and $H_1^0$ are mixed states of the two singlets, they involve essentially the same production and decay processes.  For definiteness in our analysis, we ignore the possibility of $\varphi \to \varphi' \varphi'$, with $\varphi$ denoting any of the other neutral Higgs bosons, because of the uncertainties in the triple Higgs couplings. 
Due to its odd CP property, the $H_3^0$ boson does not couple to the weak gauge bosons at tree level.  Therefore, it can only be produced via the gluon-gluon fusion (GGF) and top associated production channels.  Its decay modes do not include the weak boson pairs.  For sufficiently high mass, in particular, it can also decay into $hZ$.  On the other hand, the $H_5^0$ boson comes purely from the weak isospin triplet fields and does not couple to the SM fermions at tree level.  Consequently, it can only be produced through the vector boson fusion (VBF) and Higgs-strahlung production processes, and decay only to the final states of $WW$, $ZZ$, and $\gamma\gamma$.  It is our objectives in this work to study in detail the above-mentioned features of the neutral Higgs bosons in the GM model and to constrain the model using the available search data.

This paper is organized as follows.  Section~\ref{sec:model} briefly reviews the GM model and, in particular, provides the couplings of the neutral Higgs bosons with the SM fermions and weak gauge bosons.  We show how such couplings vary from their SM values as one adjusts the triplet VEV and the mixing angle $\alpha$ between the two singlets.  In view of good agreement between the measured Higgs data and the SM predictions, we will focus on the region of small $\alpha$.  In Section~\ref{sec:br}, we discuss the decay branching ratios and the signal strengths of different production channels for each of the neutral Higgs bosons in the model.  Section~\ref{sec:constraints} discusses the constraints on the GM model based upon experimental searches of another SM-like Higgs bosons in various channels.  This complements the other constraints from doubly-charged Higgs boson searches and indirectly from $B$ physics and electroweak precision data.  We summarize our analysis in Section~\ref{sec:summary}.

\section{Georgi-Machacek Model \label{sec:model}}

The Higgs sector in the GM model comprises an isospin doublet field $\phi$ with $Y=1/2$, 
a complex triplet field $\chi$ with $Y=1$, and a real triplet field $\xi$ with $Y=0$, 
where the electric charge $Q$ is given by $Q=T_3+Y$ with $T_3$ being the third isospin generator.
Organized in an $SU(2)_L\times SU(2)_R$ covariant form, one has:
\begin{align}
\Phi=
\begin{pmatrix}
\phi^{0*} & \phi^+ \\
-(\phi^+)^* & \phi^0
\end{pmatrix} ~,~
\Delta=
\begin{pmatrix}
\chi^{0*} & \xi^+ & \chi^{++} \\
-(\chi^+)^* & \xi^0 & \chi^{+} \\
(\chi^{++})^* & -(\xi^+)^* & \chi^{0} 
\end{pmatrix} ~,
\label{eq:Higgs_matrices}
\end{align}
where the phase convention for the component scalar fields is such that $\phi^- = (\phi^+)^*$, $\chi^{--} = (\chi^{++})^*$, $\chi^{-} = (\chi^{+})^*$, $\xi^{-} = (\xi^{+})^*$.  Moreover, the neutral components after electroweak symmetry breaking are parameterized as 
\begin{align}
\phi^0 = \frac{1}{\sqrt{2}}(v_\phi+\phi_r+i\phi_i) ~,~ 
\chi^0 = v_\chi+\frac{1}{\sqrt{2}}(\chi_r+i\chi_i) ~,~
\xi^0 = v_\xi+\xi_r ~, \label{eq:neutral}
\end{align}
where $v_\phi$, $v_\chi$ and $v_\xi$ denote the VEV's of $\phi$, $\chi$ and $\xi$, respectively.  
The most general Higgs potential consistent with the $SU(2)_L\times SU(2)_R\times U(1)_Y$ symmetry has four dimensionful and five dimensionless parameters.   Its explicit form can be found, for example, in Ref.~\cite{Chiang:2012cn}.

When the VEV's of the two triplet fields are aligned: $v_\chi=v_\xi \equiv v_3$, the original $SU(2)_L\times SU(2)_R$ symmetry in the Higgs potential of the model breaks down to the custodial $SU(2)_V$ symmetry.  
In this case, the masses of the $W$ and $Z$ bosons have exactly the same form as in the SM: $M_W^2 = g^2v^2 / 4$ and $M_Z^2=g^2v^2 / (4\cos^2\theta_W)$, where $\theta_W$ is the weak mixing angle and $v^2\equiv v_\phi^2 + 8v_3^2 = (246~{\rm GeV})^2$.  Therefore, the electroweak $\rho$ parameter keeps unity at tree level.  Following the convention in Ref.~\cite{Kanemura:2014bqa}, we define the ratio of the VEV's as
\begin{align}
\tan\beta = \frac{v_\phi}{2\sqrt{2} v_3} ~.
\end{align}
Note that $\tan\beta$ is the reciprocal of $\tan\theta_H$ used in most other works and goes to infinity in the SM limit.

The component scalar fields can be classified into irreducible representations of $SU(2)_V$ multiplets: 
the $\Phi$ field is decomposed as ${\bf 2}\otimes {\bf 2} \to {\bf 3}\oplus{\bf 1}$, and the $\Delta$ field as ${\bf 3}\otimes {\bf 3} \to {\bf 5}\oplus{\bf 3}\oplus{\bf 1}$ after the $SU(2)_L \otimes SU(2)_R \to SU(2)_V$ symmetry breaking.  As a result, we have one 5-plet, two 3-plet and two singlet representations under the custodial $SU(2)_V$ symmetry.  
The 5-plet, denoted by $H_5 = (H_5^{\pm\pm},H_5^\pm,H_5^0)^T$, arises within the $\Delta$ field.  The two 3-plet fields mix through the angle $\beta$ to render a physical CP-odd Higgs 3-plet, denoted by $H_3 = (H_3^\pm,H_3^0)^T$, and another NG 3-plet, $(G^\pm,G^0)^T$, to become the longitudinal components of the weak gauge bosons.  The two CP-even singlet fields further mix by an angle $\alpha$, determined by the quartic coupling constants in the Higgs potential, to produce the SM-like Higgs boson $h$ and another physical singlet denoted by $H_1^0$.  Due to the custodial symmetry, Higgs bosons belonging to the same $SU(2)_V$ multiplet are degenerate in mass.  Therefore, we will simply use $M_{H_{1,3,5}}$ to denote the masses of $H_{1,3,5}^0$, respectively.  For later uses, we write out the physical neutral Higgs fields in terms of the original component scalar fields as follows:
\begin{align}
\begin{pmatrix}
h \\ H_1^0 \\ H_5^0
\end{pmatrix}
&= 
\begin{pmatrix}
\cos\alpha & -\sin\alpha & 0 \\
\sin\alpha & \cos\alpha & 0 \\
0 & 0 & 1
\end{pmatrix}
\begin{pmatrix}
1 & 0 & 0 \\
0 & \sqrt{\frac13} & \sqrt{\frac23} \\
0 & -\sqrt{\frac23} & \sqrt{\frac13}
\end{pmatrix}
\begin{pmatrix}
\phi_r \\ \xi_r \\ \chi_r
\end{pmatrix} ~, \notag \\
H_3^0 
&= \sin\beta~\chi_i - \cos\beta~\phi_i ~.
\end{align}
Here we identify the quantum of the $h$ field as the 125-GeV Higgs boson found at the LHC.  
Note that, the $H_3^0$ does not couple to the weak gauge bosons at tree level, whereas the $H_5^0$ does not couple to the SM fermions.

Denote the couplings of a neutral Higgs boson $\varphi$ ($= h, H_1^0, H_3^0$ or $H_5^0$) to a SM fermion pair and a weak gauge boson in the model by $g_{\varphi ff}$ and $g_{\varphi VV}$, respectively.  Define the scaling factors $\kappa_F$ and $\kappa_V$ as
\begin{align}
\kappa_F = \frac{g_{\varphi ff}}{g_{hff}^{\rm SM}} ~,~\mbox{and}~
\kappa_V = \frac{g_{\varphi VV}}{g_{hVV}^{\rm SM}} ~,
\end{align}
where $g_{hff}^{\rm SM}$ and $g_{hVV}^{\rm SM}$ refer to the corresponding values in the SM.  The scaling factors for the neutral Higgs bosons in the GM model are summarized in Table~\ref{tab:couplings}.  Since $H_3$ is CP-odd, its coupling with fermions involves a $\gamma_5$ factor that is not reflected in the scaling factor given in the table.

\begin{table}[th]
\begin{tabular}{c|cc}
\hline\hline
Higgs boson & ~~~~~~~~~~$\kappa_F$~~~~~~~~~~ & ~~~~~~~~~~$\kappa_V$~~~~~~~~~~ \\
\hline
$h$ & $\cos\alpha / \sin\beta$ & $\sin\beta\cos\alpha - \sqrt{8/3} \cos\beta\sin\alpha$
\\
$H_1^0$ & $\sin\alpha / \sin\beta$ & $\sin\beta\sin\alpha + \sqrt{8/3} \cos\beta\cos\alpha$
\\
$H_3^0$ & $i\eta_f\cot\beta$ & 0
\\
$H_5^0$ & 0 & $\kappa_W = -\cos\beta / \sqrt{3}$ and $\kappa_Z = 2\cos\beta / \sqrt{3}$
\\
\hline\hline
\end{tabular}
\vspace{6pt}
\caption{Scaling factors of the couplings between neutral Higgs bosons in the GM model with the SM fermions and weak gauge bosons.  $\eta_f = +1$ for up-type quarks and $-1$ for down-type quarks and charged leptons.}
\label{tab:couplings}
\end{table}

\begin{figure}[th]
\centering 
\includegraphics[width=7cm]{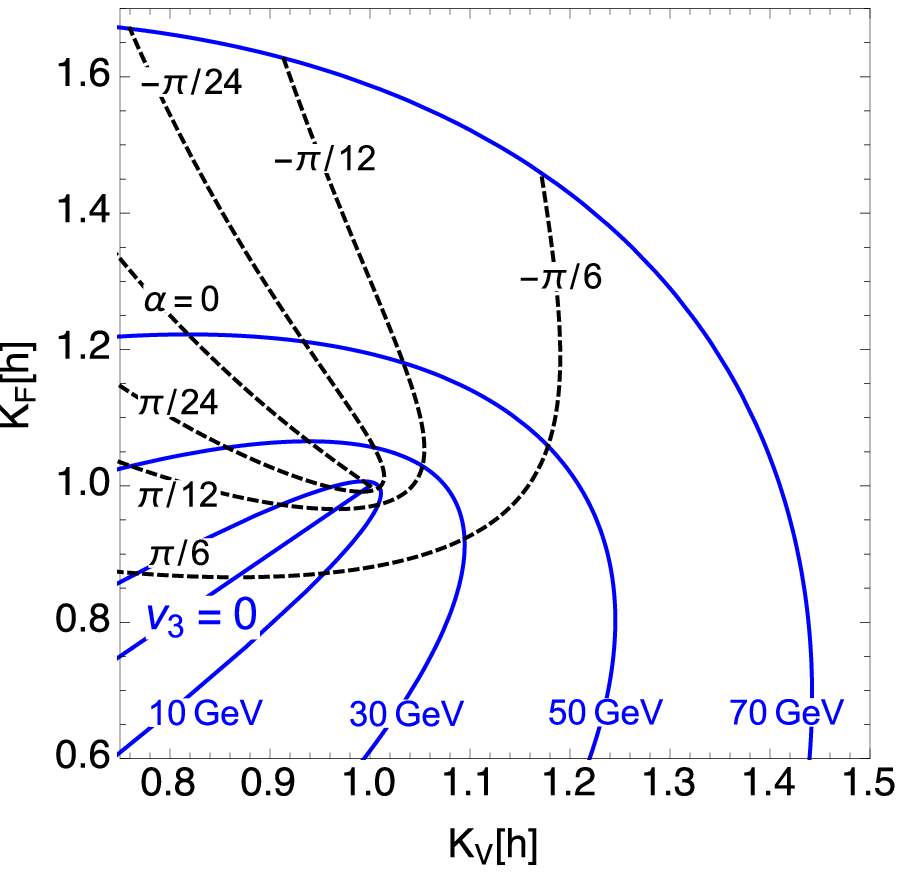}
\hspace{0.5cm}
\includegraphics[width=7cm]{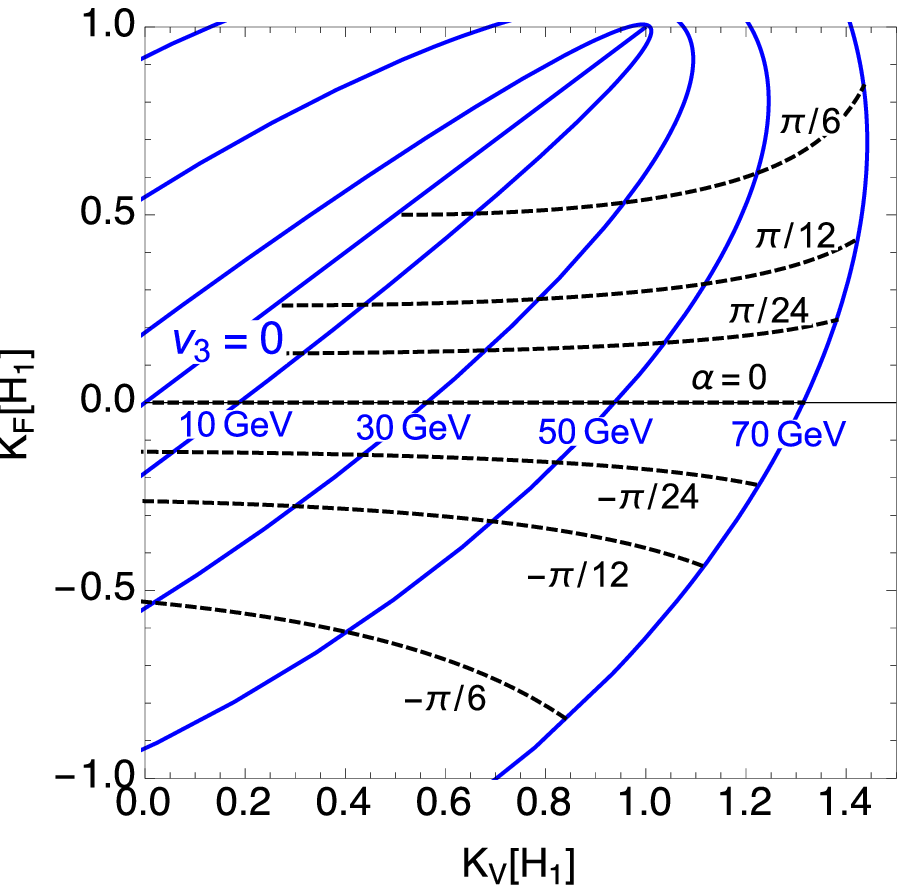}
\\
\vspace{-0.3cm}
(a) \hspace{7cm} (b)
\\
\bigskip
\includegraphics[width=7cm]{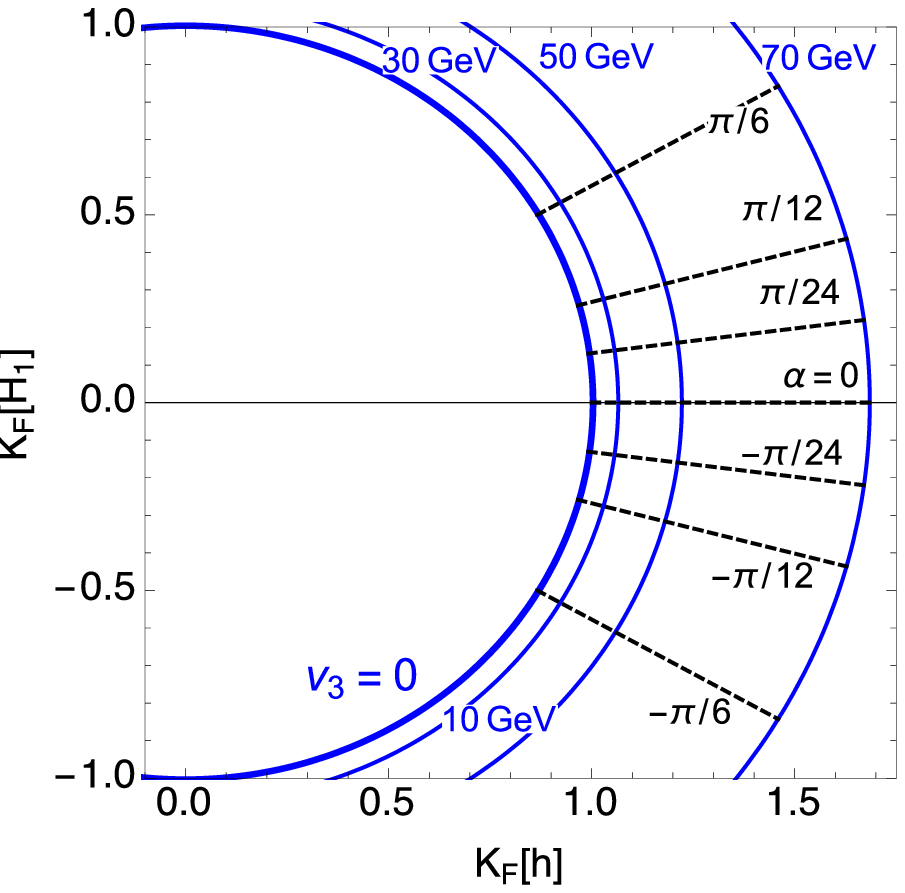}
\hspace{0.5cm}
\includegraphics[width=7cm]{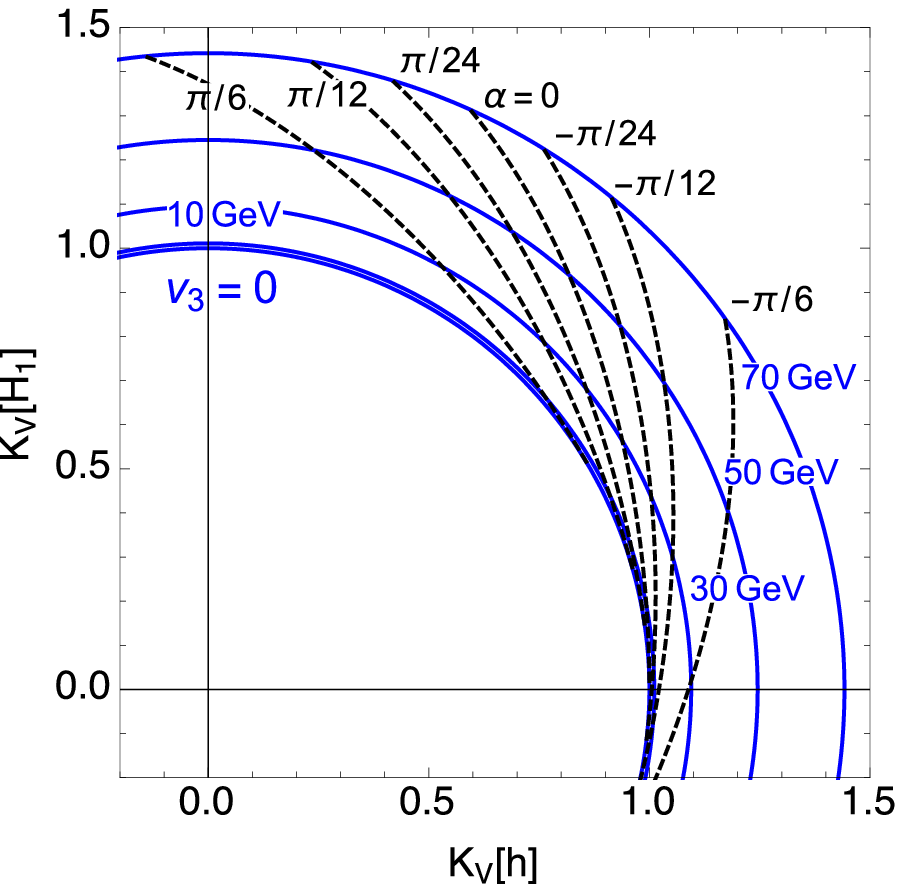}
\\
\vspace{-0.3cm}
(c) \hspace{7cm} (d)
\caption{Scaling factors for the two custodial singlets, $h$ and $H_1^0$.  The blue solid contours are for $v_3 = 0, 10, 30, 50$, and $70$ GeV with varying $\alpha$.  The black dotted curves are for $\alpha = -\pi/6, -\pi/12, -\pi/24, 0, \pi/24, \pi/12$, and $\pi/6$ with varying $\beta$.
}
\label{FIG:ScalingFactors}
\end{figure}

As only the two custodial singlets in the model can possibly couple to both SM fermions and weak bosons simultaneously, we plot in Fig.~\ref{FIG:ScalingFactors} the contours of different pairs of scaling factors for the two custodial singlets.  In Figs.~\ref{FIG:ScalingFactors}(a) and \ref{FIG:ScalingFactors}(b), contours are shown on the $\kappa_V$-$\kappa_F$ plane for $h$ and $H_1^0$, respectively, for various representative values of $v_3$ and $\alpha$.  In fact, the blue solid contours in both plots are identical as the scaling factors of $h$ for the phase $\alpha$ are the same as those of $H_1^0$ for the phase $\alpha + \pi/2$.  This also explains that the fixed-$\alpha$ contours (black dotted ones) have the $\alpha \leftrightarrow \alpha + \pi/2$ correspondence between the two plots.  Similarly, Fig.~\ref{FIG:ScalingFactors}(c) shows contours on the $\kappa_F^h$-$\kappa_F^{H_1}$ plane, and Fig.~\ref{FIG:ScalingFactors}(d) those on the $\kappa_V^h$-$\kappa_V^{H_1}$ plane.  The blue solid contours of fixed $v_3$ values in the two plots depict a series of circles concentric at the origin, of radii $1/\sin\beta$ and $\sqrt{1 + (5/3)\cos^2\beta}$, respectively.  The only difference is that the black dotted contours of fixed $\alpha$ values are along the radial direction in the former, but spiral out in the latter.

Note that in the limit $\alpha = 0$, meaning that the singlet from the $\Phi$ field does not mix with that from the $\Delta$ field, the scaling factors $(\kappa_F,\kappa_V) = (1/\sin\beta,\sin\beta)$ for $h$ and $(0,\sqrt{8/3} \cos\beta)$ for $H_1^0$.  Therefore, the two scaling factors of $h$ are reciprocal to each other with $\kappa_F \ge 1$ and $\kappa_V \le 1$, as indicated by the $\alpha = 0$ contours in Fig.~\ref{FIG:ScalingFactors}(a).  The $H_1^0$ boson becomes fermiophobic but still allows a nonzero coupling with the weak gauge bosons, depending solely upon the value of $v_3$, as shown by the horizontal dotted line in Fig.~\ref{FIG:ScalingFactors}(b).  In contrast, the couplings of $H_5^0$ to the weak bosons are purely proportional to $v_3$, independent of $\alpha$.  $H_3^0$ also has couplings with the SM fermions as a result of mixing with the custodial triplet in the $\Phi$ field.

On the other hand, $\sin\beta \to 1$ in the limit of vanishing triplet VEV, $v_3 = 0$.  In this case, $(\kappa_F,\kappa_V) = (\cos\alpha,\cos\alpha)$ for $h$ and $(\sin\alpha,\sin\alpha)$ for $H_1^0$, shown as the two blue diagonal lines in Figs.~\ref{FIG:ScalingFactors}(a) and \ref{FIG:ScalingFactors}(b), while both $H_3^0$ and $H_5^0$ totally decouple from the SM fermions and weak bosons.

\section{Branching Ratios and Signal Strengths \label{sec:br}}

We discuss in this section patterns of the branching ratios and signal strengths for each of the neutral Higgs bosons in the model.  In this paper, we define the signal strength of a particular Higgs production channel at LHC as the production rate in the GM model normalized to the SM production rate for a fictitious Higgs with the same mass:
\begin{align}
\mu_{X}[\varphi] = 
\frac{\sigma^{\rm GM}(pp \to \varphi){\cal B}^{\rm GM}(\varphi \to X)}
     {\sigma^{\rm SM}(pp \to \varphi){\cal B}^{\rm SM}(\varphi \to X)} ~,
\end{align}
where $X$ is some decay final state of $\varphi$, $\sigma$ and $\cal B$ denote respectively the cross section and branching ratio, and the narrow width approximation for $\varphi$ has been assumed.

\subsection{The $h$ boson \label{sec:h}}

\begin{figure}[th]
\centering 
\includegraphics[width=7cm]{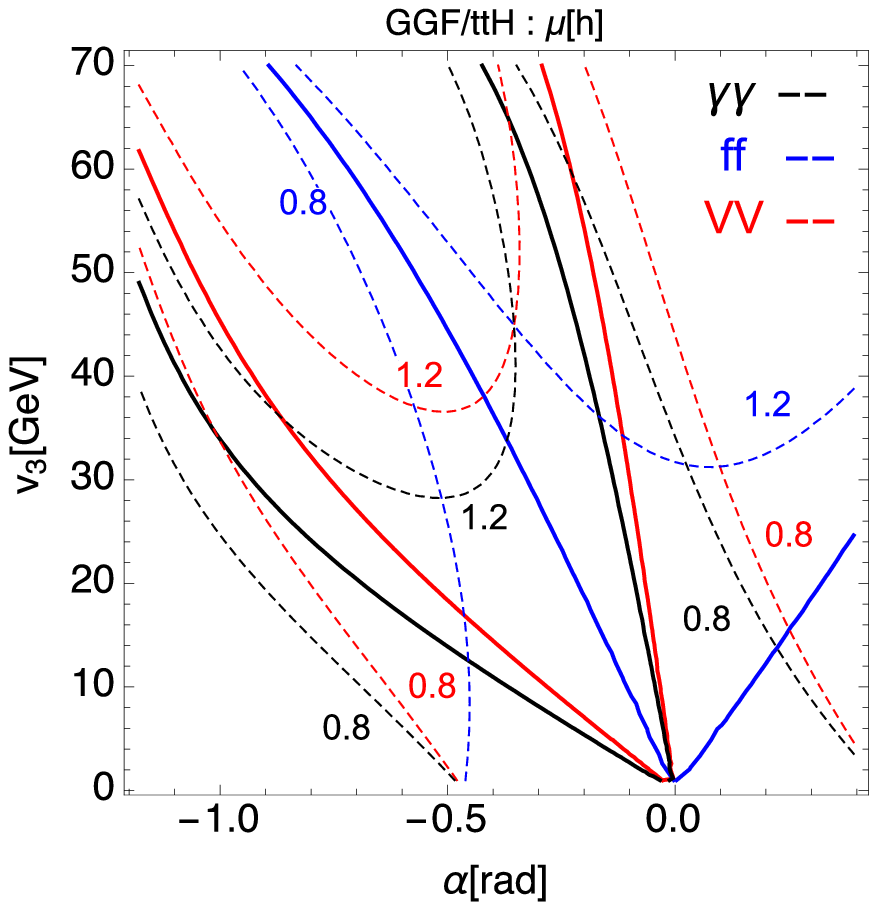}
\hspace{0.5cm}
\includegraphics[width=7cm]{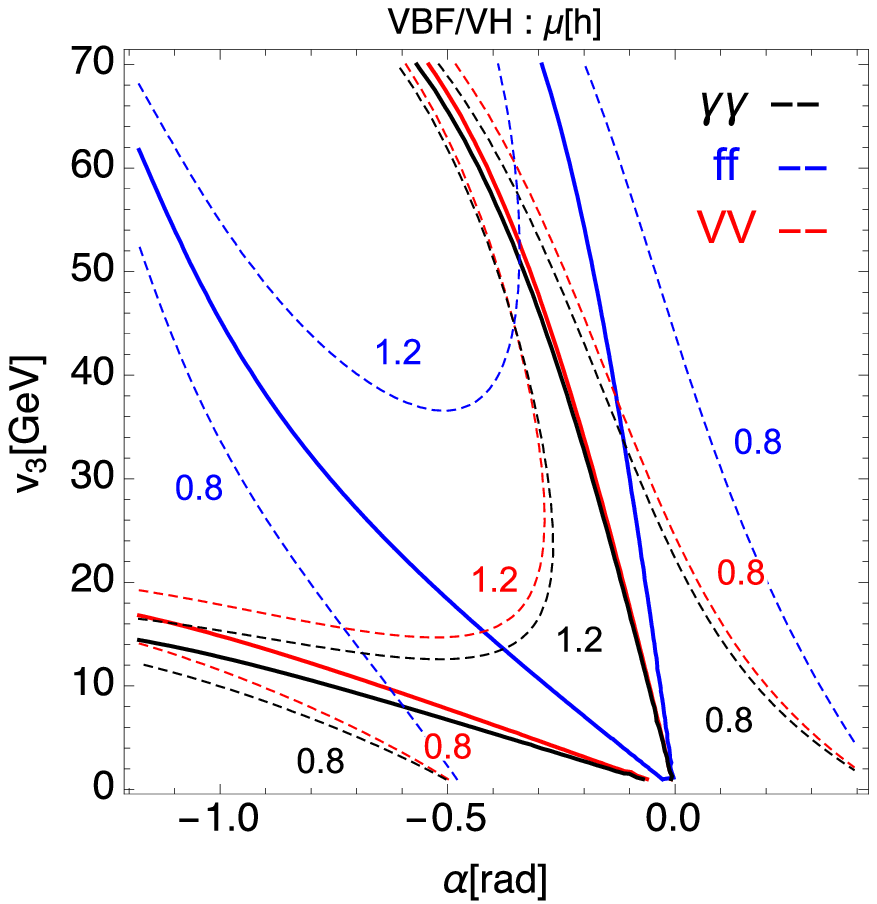}
\\
\vspace{-0.3cm}
(a) \hspace{7cm} (b)
\caption{Signal strength contours for the $h$ boson in (a) the GGF/ttH production channels and (b) the VBF/VH production channels.  The solid curves correspond to unit signal strengths, and the dotted curves are directly labeled with the signal strength values.}
\label{FIG:SignalStrengthSM}
\end{figure}

Since the SM-like Higgs boson $h$ generally has different couplings with SM particles from the SM expectation, the signal strengths of various search channels can be different from unity.  Therefore, one could use the measured signal strengths to constrain the parameters $\alpha$ and $\beta$, as have been done in Ref.~\cite{Chiang:2013rua}.  Instead of performing another global fit as in Ref.~\cite{Chiang:2013rua}, here we want to study the effects of individual production channels.  In Fig.~\ref{FIG:SignalStrengthSM}, we plot the signal strength contours for the $h$ boson in the model.  Plot~(a) shows those for the channels of $\gamma\gamma$, $f\bar f$, and $VV^*$ via the GGF and top associated (ttH) productions, and plot~(b) for the same channels via the VBF and Higgs-strahlung (VH) productions.  In both plots, we depict the contours for each signal strength $\mu = 1$ by solid curves, as well as $\mu = 0.8$ and $1.2$ by dotted curves.  We note in passing that we have ignored the effects of charged Higgs bosons in the $h \to \gamma\gamma$ decays.  This is justified in view of the current overall agreement of signal strengths with the SM expectations.

Because of higher statistics, the signal strengths of $ZZ^*(\to 4\ell, 2\ell \, 2\nu)$ and $\gamma\gamma$ channels in the GGF production are measured more accurately than in the VBF production.  The latter production channels should have a much better precision at the high-luminosity LHC (HL-LHC).  Due to better detector sensitivity, these bosonic channels are superior to the fermionic decay channels ($bb$ and $\tau\tau$).  The $\tau\tau$ mode in the VBF production has a relatively better sensitivity among the fermionic decays.

As shown in Fig.~\ref{FIG:SignalStrengthSM}, the signal strengths of the bosonic channels for non-negligible $v_3$ have a strong dependence on $\alpha$.  Therefore, one can easily use their current values to exclude the parameter space of $\alpha \agt +0.5$.  Moreover, the region for each $0.8 \le \mu \le 1.2$ have two branches at large $v_3$.  Thanks to the different shapes in different channels in both GGF and VBF productions, the region with $\alpha \alt -0.5$ in one of the branches can also be excluded. 
The remaining parameter space can be constrained effectively once we are able to measure the $ff$ decay channels in the GGF process with sufficiently good precision, particularly the $\tau\tau$ channel in the 1-jet category.
In addition, if the signal strengths of the $\gamma\gamma$ and $VV$ channels in the VBF and GGF processes can be separately measured to a good accuracy, it is also possible to constrain the remaining parameter space, especially in the large $v_3$ region.

In summary, the current Higgs data point us to the region of small mixing angle $\alpha$.  Therefore, we will focus our attention to such cases in the following analyses.  More explicitly, we will consider the examples of $\alpha = -\pi/24$ and $-\pi/12$ with $v_3 = 10$, $30$, and $70$ GeV.

\subsection{The $H_1^0$ boson \label{sec:H1}}

\begin{figure}[th]
\centering 
\includegraphics[width=5cm]{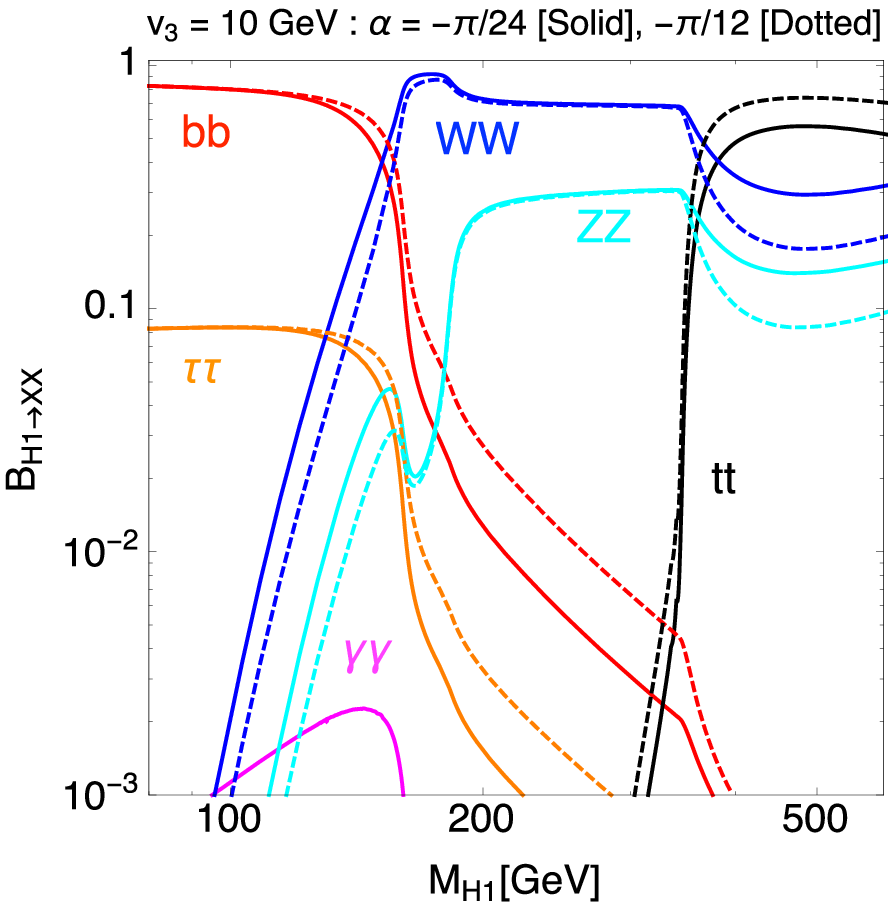}
\hspace{0.25cm}
\includegraphics[width=5cm]{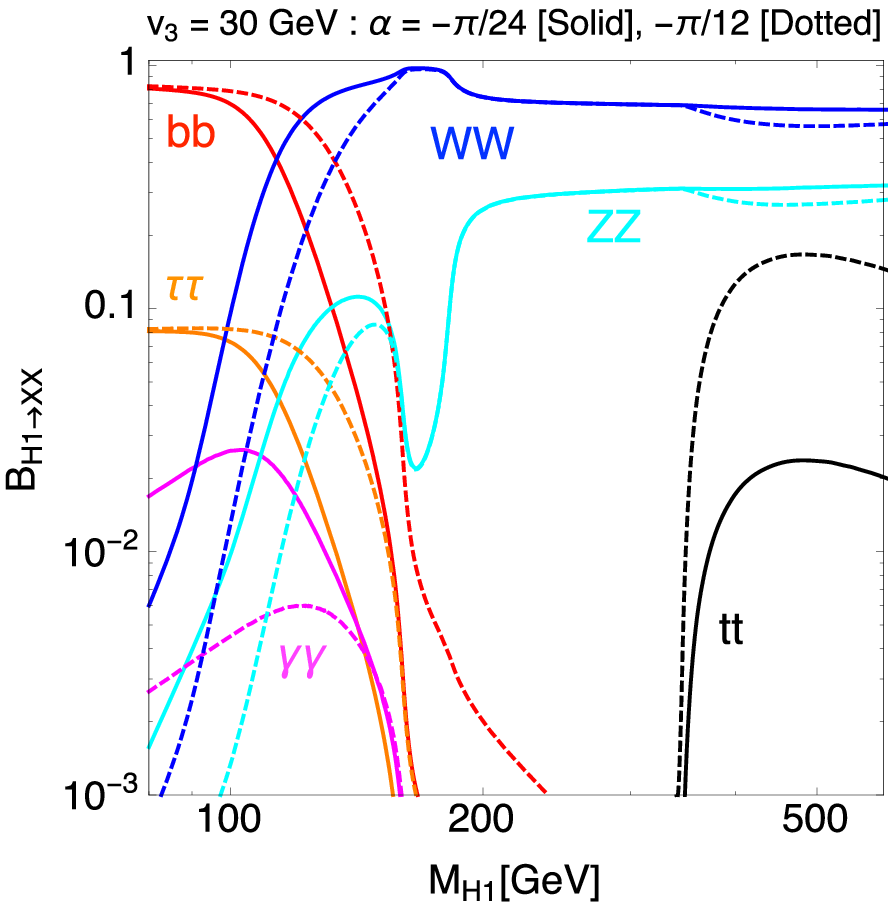}
\hspace{0.25cm}
\includegraphics[width=5cm]{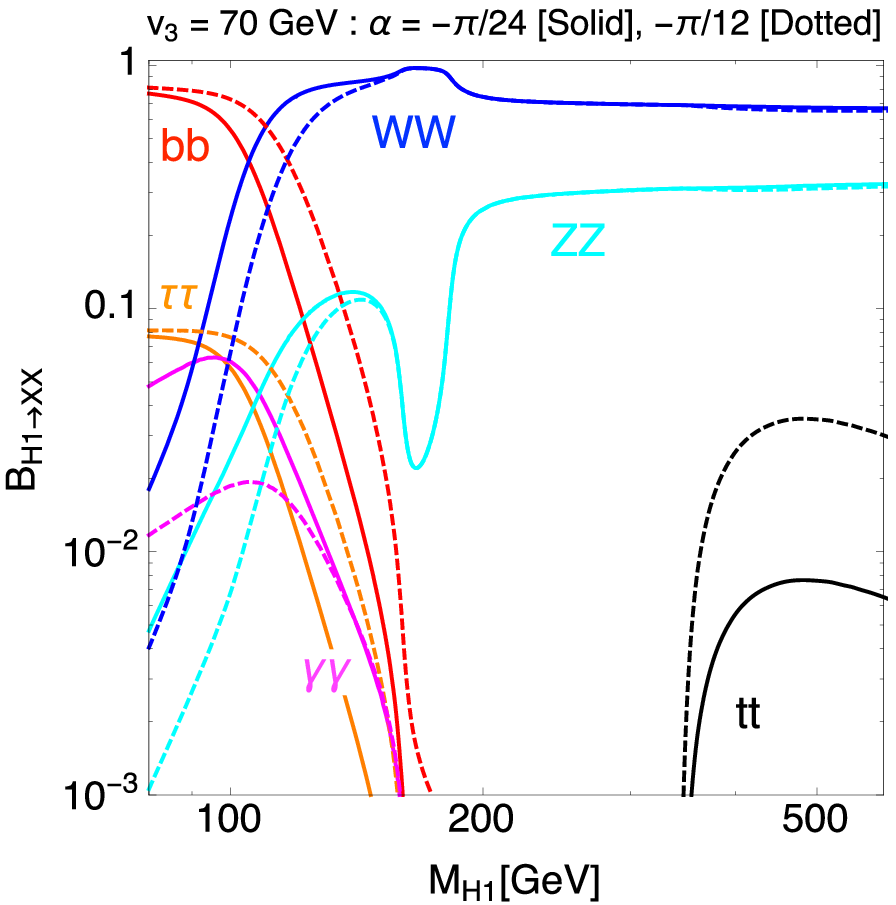}
\\
\vspace{-0.3cm}
(a) \hspace{4.75cm} (b) \hspace{4.75cm} (c)
\caption{Decay branching ratios of the $H_1^0$ boson as functions of $M_{H_1}$.  We take (a) $v_3=10$ GeV, (b) $30$ GeV, and (c) $70$ GeV.  
}
\label{FIG:BRH1}
\end{figure}

The decay branching ratios of the other custodial singlet, the $H_1^0$ boson, are shown in Fig.~\ref{FIG:BRH1} for $v_3=10$ GeV, $30$ GeV and $70$ GeV.  The curves are calculated using the data provided by the LHC Higgs Cross Section Working Group~\cite{Ref:XSWG} with the corresponding correction factors.  For the loop-induced processes, leading-order (one-loop) correction factors are evaluated.
In each plot, we draw curves for $\alpha = -\pi/24$ (solid) and $-\pi/12$ (dotted).  A comparison between the solid and dotted curves reveal the fermiophobic nature of the $H_1^0$ boson for small $\alpha$, as alluded to earlier.

When $v_3 = 10$ GeV or smaller, the dominant decay mode is $b\bar b$ when the $H_1^0$ boson mass $M_{H_1} \alt 2M_W$ and switches to the vector boson pairs when $M_{H_1} \agt 2M_W$.  This feature is the same as a SM-like Higgs boson with a varying mass.  The transition of dominant mode happens at a lower mass for larger $v_3$.  It is also observed that the decay branching ratio of the $\gamma\gamma$ mode in the low $M_{H_1}$ regime increases with $v_3$ due to less destructively interfering loop processes.  For sufficiently large $v_3$, the dominant decay modes are the weak gauge boson pairs.  Again, for definiteness, the charged Higgs contributions to $\gamma\gamma$ decay are not included in the calculation.  The $gg$ and $cc$ decay modes are not shown in the plots.

We also note that the $H_1^0 \to hh$ decay mode is not included in the calculation, because the relevant coupling, $\lambda_{hhH_1}^{}$, is not determined without explicitly specifying the Higgs sector parameters.  Qualitatively, a nonzero $\lambda_{hhH_1}^{}$ coupling will result in a suppression of the regular search channels when $M_{H_1} \agt 250$ GeV, and may lead to more production of four-body final states.  Owing to its small rate, the $\gamma Z$ decay channel is totally neglected here for simplicity.

\begin{figure}[th]
\centering 
\includegraphics[width=5cm]{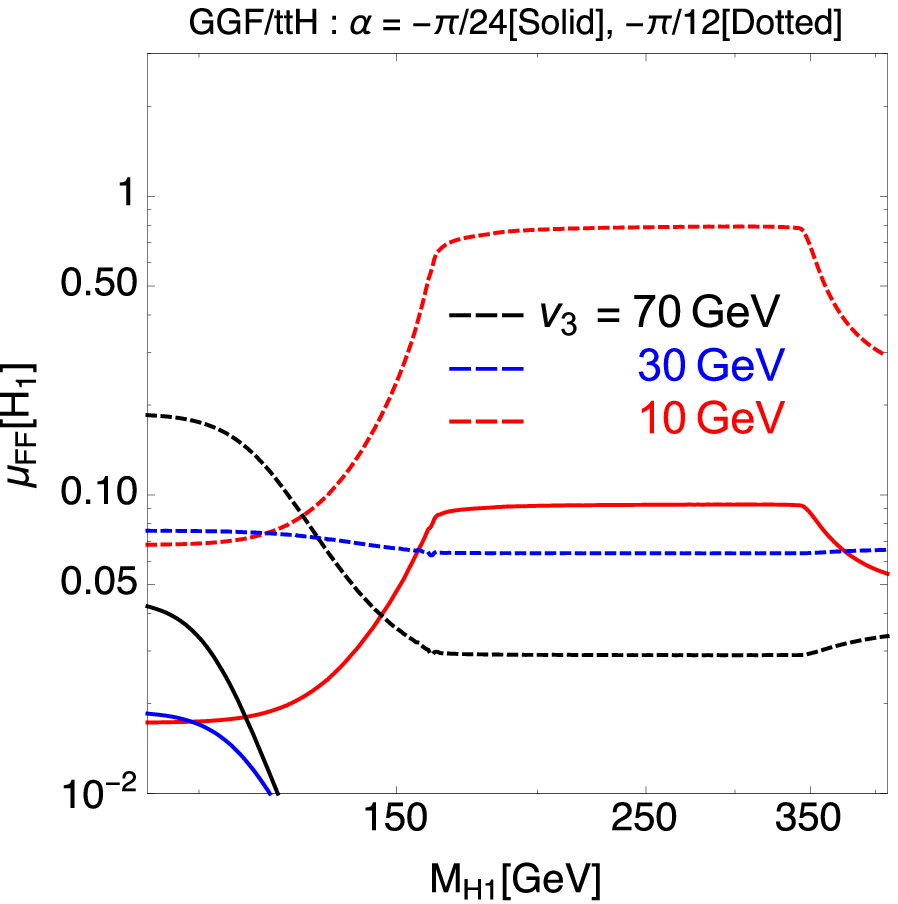}
\hspace{0.25cm}
\includegraphics[width=5cm]{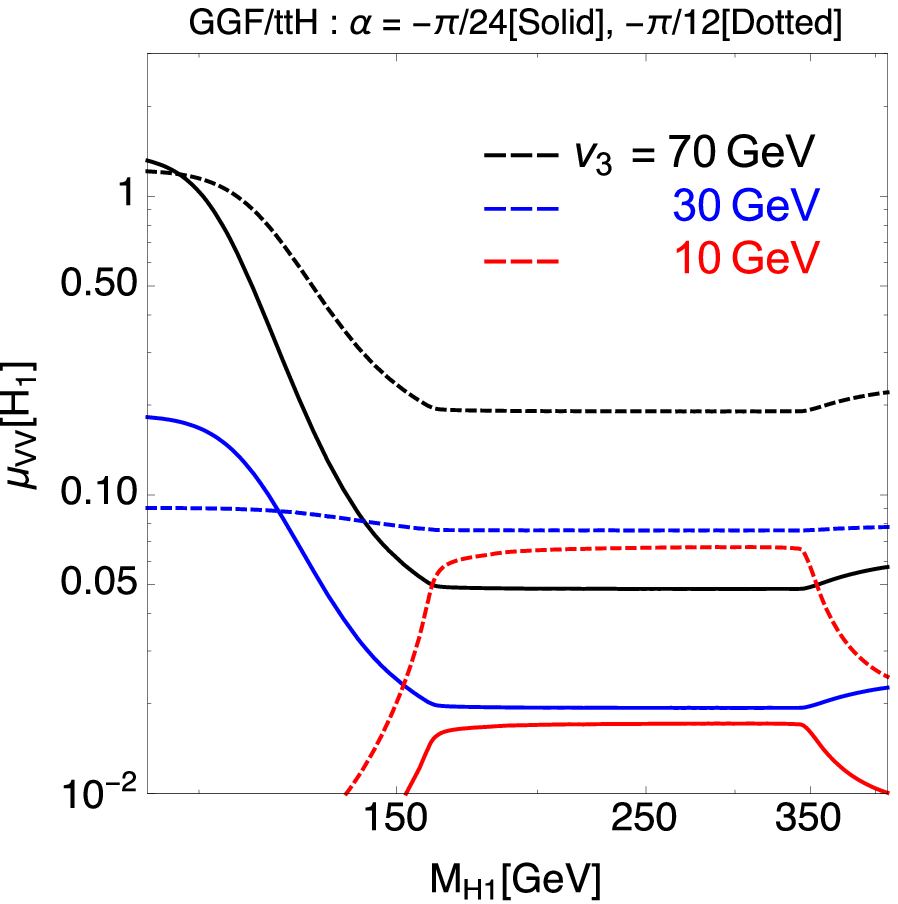}
\hspace{0.25cm}
\includegraphics[width=5cm]{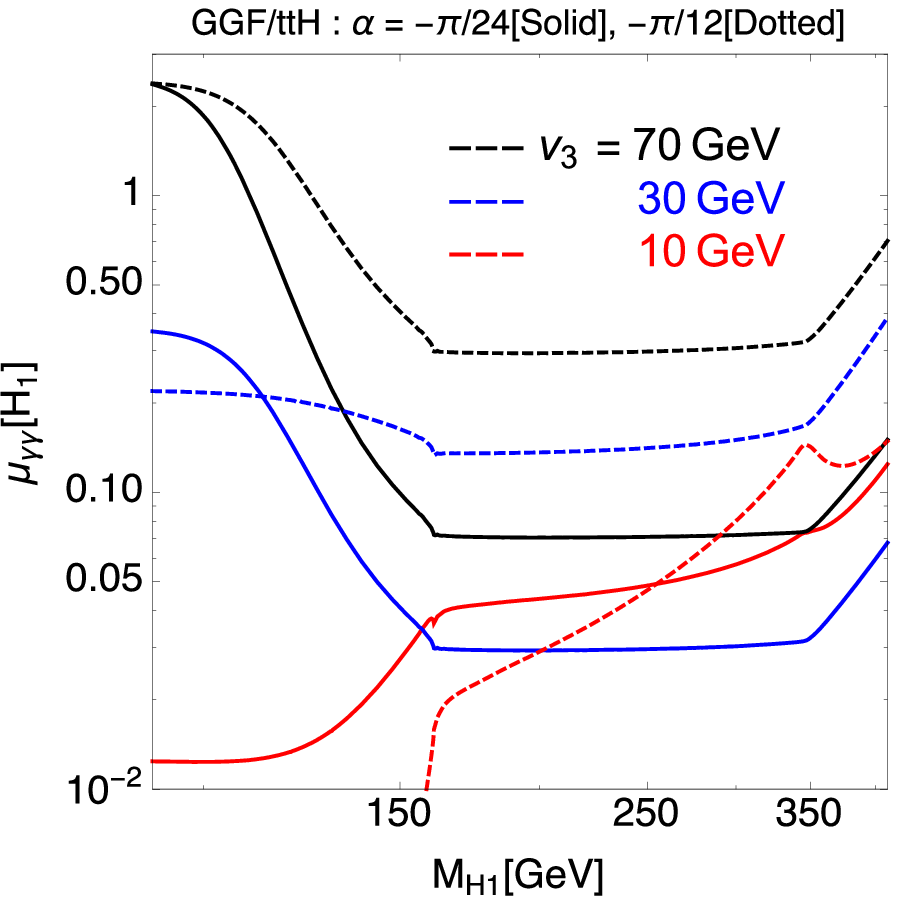}
\\
\vspace{-0.3cm}
(a) \hspace{4.75cm} (b) \hspace{4.75cm} (c)
\\
\bigskip
\includegraphics[width=5cm]{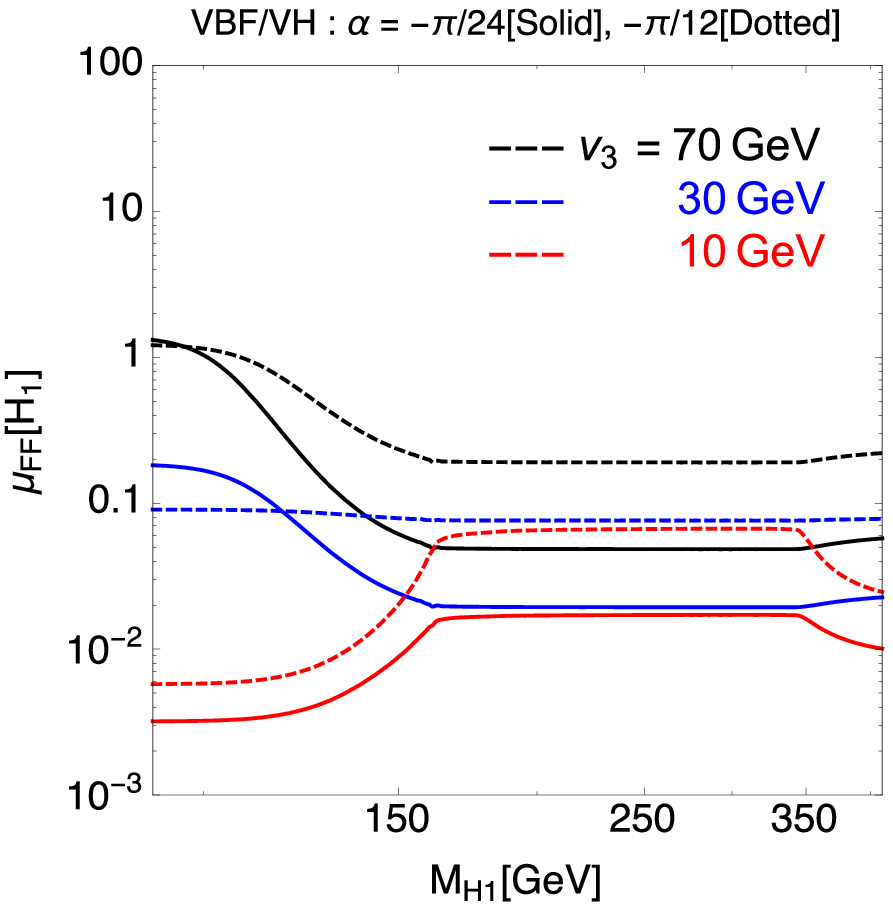}
\hspace{0.25cm}
\includegraphics[width=5cm]{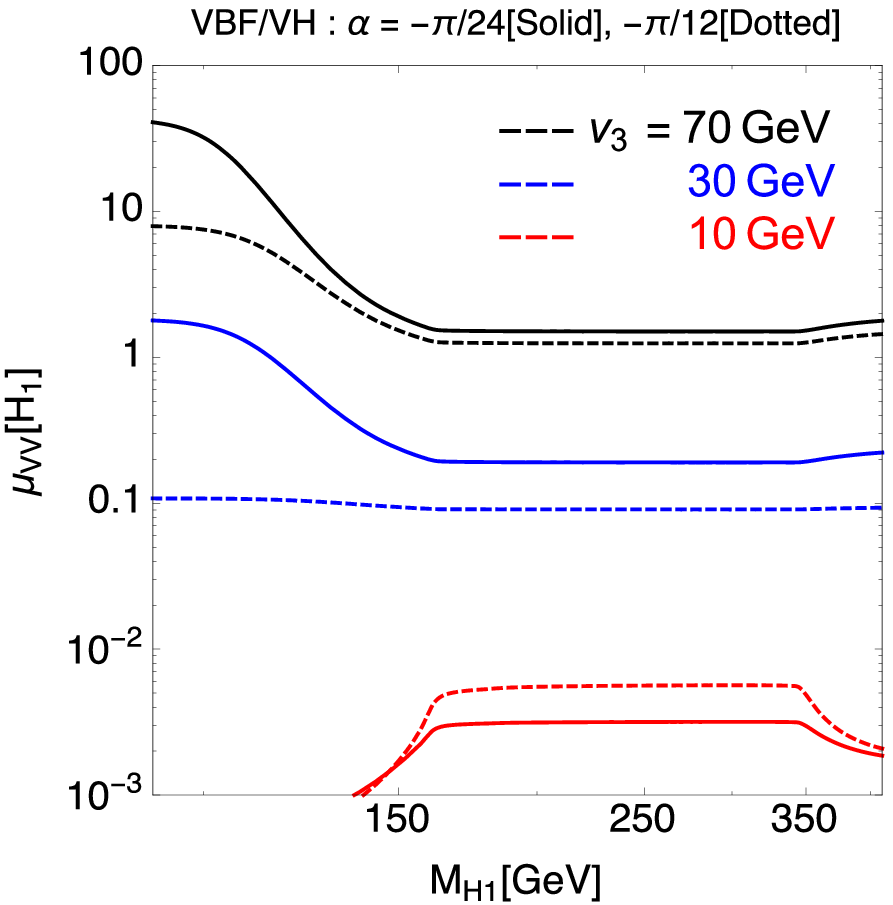}
\hspace{0.25cm}
\includegraphics[width=5cm]{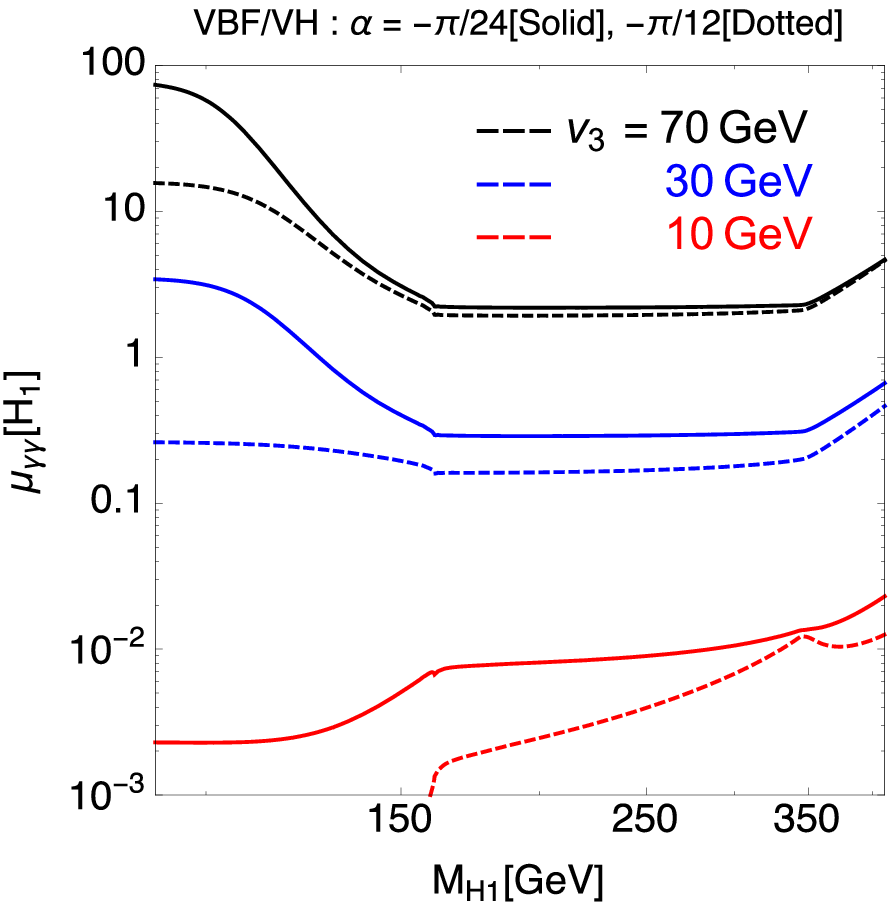}
\\
\vspace{-0.3cm}
(d) \hspace{4.75cm} (e) \hspace{4.75cm} (f)
\caption{Signal strengths of the $H_1^0$ boson as functions of $M_{H_1}$.  Plots in the left, middle, and right columns are for the $f\bar f$, $VV$, and $\gamma\gamma$ channels, respectively.  Plots in the upper and lower rows are for the GGF/ttH and VBF/VH productions, respectively.  
}
\label{FIG:SignalStrengthH1MH}
\end{figure}

We now consider major production channels for the $H_1^0$ boson.  We define the signal strength for each channel by normalizing its production rate in the GM to that for the Higgs boson in the SM but with a mass equal to $M_{H_1}$.  Therefore, the signal strengths of the $X$ channel from the GGF/ttH (referred to simply as GGF) and VBF/VH (referred to simply as VBF) productions can be respectively expressed as
\begin{align}
\mu_X^\text{GGF} [H_1]
&=
(\kappa_F^{H_1})^2 \times
\frac{{\mathcal B}_X}{{\mathcal B}_X^\text{SM}(M_{H_1})} 
\simeq 
\frac{(\kappa_F^{H_1})^2(\kappa_X^{H_1})^2}
{(\kappa_V^{H_1})^2{\mathcal B}_V^\text{SM}(M_{H_1})
+ (\kappa_F^{H_1})^2{\mathcal B}_F^\text{SM}(M_{H_1})} ~,
\label{eq:kappa_GGF}
\\
\mu_X^\text{VBF} [H_1]
&=
(\kappa_V^{H_1})^2 \times
\frac{{\mathcal B}_X}{{\mathcal B}_X^\text{SM}(M_{H_1})} 
\simeq 
\frac{(\kappa_V^{H_1})^2(\kappa_X^{H_1})^2}
{(\kappa_V^{H_1})^2{\mathcal B}_V^\text{SM}(M_{H_1})
+ (\kappa_F^{H_1})^2{\mathcal B}_F^\text{SM}(M_{H_1})} ~,
\label{eq:kappa_VBF}
\end{align}
where ${\mathcal B}_V^\text{SM}(M_{H_1})$ and ${\mathcal B}_F^\text{SM}(M_{H_1})$ denote the inclusive branching ratios of a SM Higgs boson of mass $M_{H_1}$ decaying into vector boson pairs and fermion pairs, respectively, and all the other modes ({\it e.g.}, multi-particle and $Z\gamma$ final states) are neglected in the last expressions.

The signal strengths of $H_1^0 \to f\bar f$, $VV$, and $\gamma \gamma$ are drawn respectively in the left, middle, and right columns of Fig.~\ref{FIG:SignalStrengthH1MH}.  For each channel, we further divide the plots according to the production mechanism to the GGF process (upper row) and the VBF process (lower row).

For small $\alpha$, such as $-\pi/24$ and $-\pi/12$ in our examples, the scaling factor of Yukawa couplings with $H_1^0$ is proportional to $\sin \alpha \sim \alpha$.  In this scenario, the denominators on the right-hand side of Eqs.~(\ref{eq:kappa_GGF}) and (\ref{eq:kappa_VBF}) can be dominated by the inclusive bosonic branching ratio.  This is the situation when $v_3$ is sufficiently large, as seen in Fig.~\ref{FIG:BRH1}.  This leads to $\mu_X^\text{VBF}[H_1] \sim \left(\kappa_X^{H_1}\right)^2 / {\mathcal B}_V^\text{SM}(M_{H_1})$ for large $v_3$.  The $1 / {\mathcal B}_V^\text{SM}(M_{H_1})$ factor is always an enhancement factor that is large when $M_{H_1} \alt 2M_W$ and the bosonic decay modes are subdominant, and gets close to unity when $M_{H_1} \agt 2M_W$.  This is seen in Fig.~\ref{FIG:SignalStrengthH1MH} by the decline in the curves for $v_3 = 30$ and $70$ GeV.  
In contrast, the signal strengths through the GGF process $\mu_X^\text{GGF}[H_1]$ is scaled from $\mu_X^\text{VBF}[H_1]$ by the factor $\left( {\kappa_F^{H_1}}/{\kappa_V^{H_1}} \right)^2$, which is smaller for larger $v_3$ values.  Therefore, one observes this general trend in the plots except when $v_3 = 10$ GeV.

In the $M_{H_1} \agt 2 M_W$ region, the predicted signal strengths are relatively flat until the $t\bar t$ channel opens up.  This reflects the fact that the production rates vary in the same way as the SM Higgs boson in the higher mass region.  Finally, it should be noted that the possibility of having the $H_1^0 \to hh$ decay is not considered here to avoid the uncertainty in the $\lambda_{hhH_1}$ coupling.

As we showed in Fig.~\ref{FIG:BRH1}, the $\gamma\gamma$ decay mode is important only in low mass region. 
In the right column, we find enhancements in the signal strength of $\gamma\gamma$ 
decay for both the VBF and GGF production channels. 
This enhancement is caused by the kinematical (off-shell) reduction in the $VV$ decay channels, suppressed fermionic decays by the $v_3/v_\phi$ factor, and less destructive interference between the $W$ and top loops.  
The enhancement is substantial for the VBF process, as the GGF process is suppressed by the fermionic scaling factor $\kappa_F^{H_1}$.

Because the $H_1^0$ decay branching fractions have relatively stable predictions for $M_{H_1} \agt 2 M_Z$, let's focus for the moment on this mass regime as the corresponding signal strengths also give stable predictions. 
The $VV$ decay channels are important once their on-shell decays become open.  A significant fraction of $H_1^0$ decays into $VV$, $\sim 60\, (\sim30) \%$ for the $WW (ZZ)$ mode, similarly to the branching ratios of a high-mass SM Higgs boson.

\begin{figure}[th]
\centering 
\includegraphics[width=7cm]{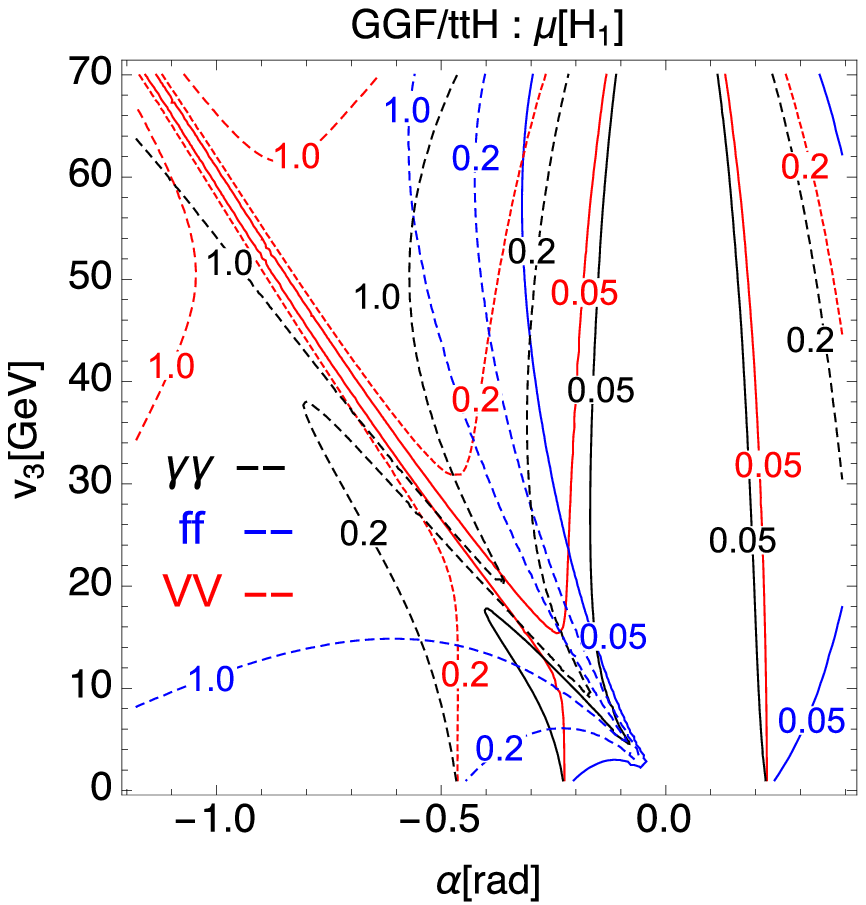}
\hspace{0.5cm}
\includegraphics[width=7cm]{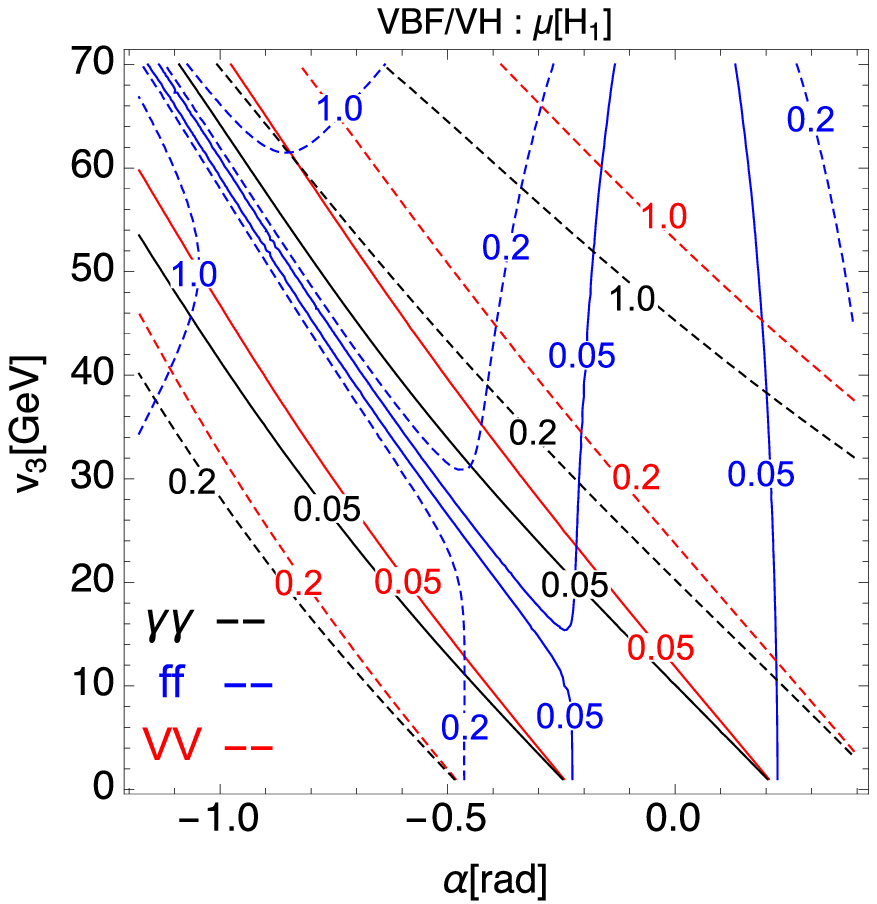}
\\
\vspace{-0.3cm}
(a) \hspace{7cm} (b)
\caption{Contours of signal strengths for a high-mass $H_1^0$ boson in (a) the GGF production and (b) the VBF production shown on the $\alpha$-$v_3$ plane.}
\label{FIG:SignalStrengthH1}
\end{figure}

Fig.~\ref{FIG:SignalStrengthH1} shows the signal strengths for various channels of $H_1^0$.  Plot (a) involves the channels through the GGF production, and plot (b) those through the VBF production.
The signal strengths of fermionic decay modes are approximately symmetric under $\alpha \to -\alpha$ 
and vanish for $\alpha = 0$. 
The singular behaviors along the diagonal line of $v_3 \sim -(\sqrt3/8) v \alpha$ is caused by 
$\kappa_V^{H_1} \sim 0$. 
Since we are mainly interested in $|\alpha| \alt 0.5$, the fermionic decay channels have less restricting power in this part of the parameter space. 
The bosonic decay modes in the GGF process are also insensitive to parameter variation within the region. 
Considering the region $\alpha > -0.5$, the $ZZ$ and $WW$ decay channels in the VBF process have a potential to constrain the parameter space with $v_3 \agt 40$ GeV if the signal strength is found to be less than $0.05$. 
The decay branching ratio of $H_1 \to \gamma\gamma$ is smaller than $10^{-3}$, a rate that is difficult to probe in the VBF production.

\subsection{The $H_3^0$ boson \label{sec:H3}}

\begin{figure}[th]
\centering 
\includegraphics[width=5cm]{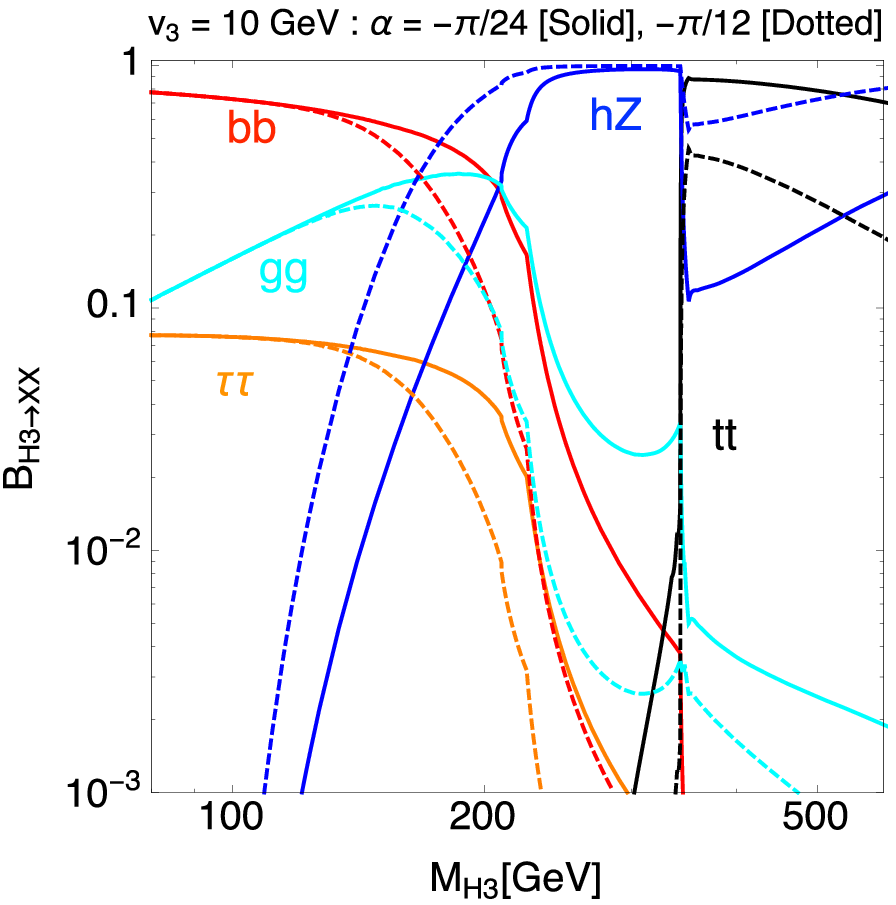}
\hspace{0.25cm}
\includegraphics[width=5cm]{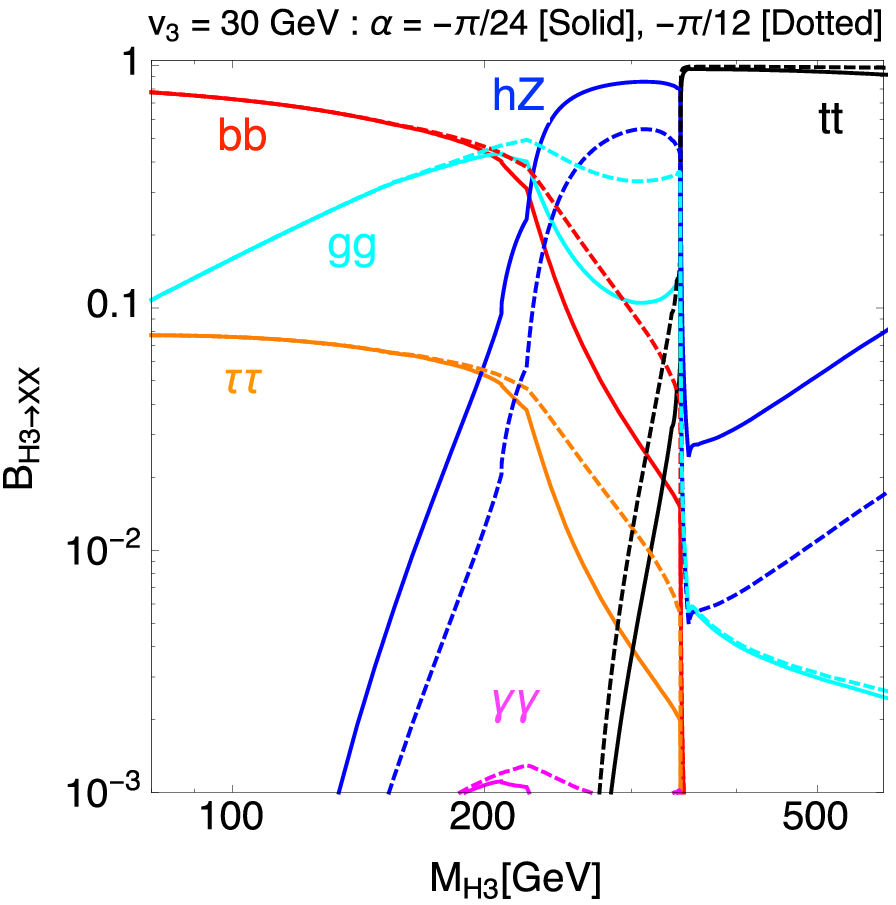}
\hspace{0.25cm}
\includegraphics[width=5cm]{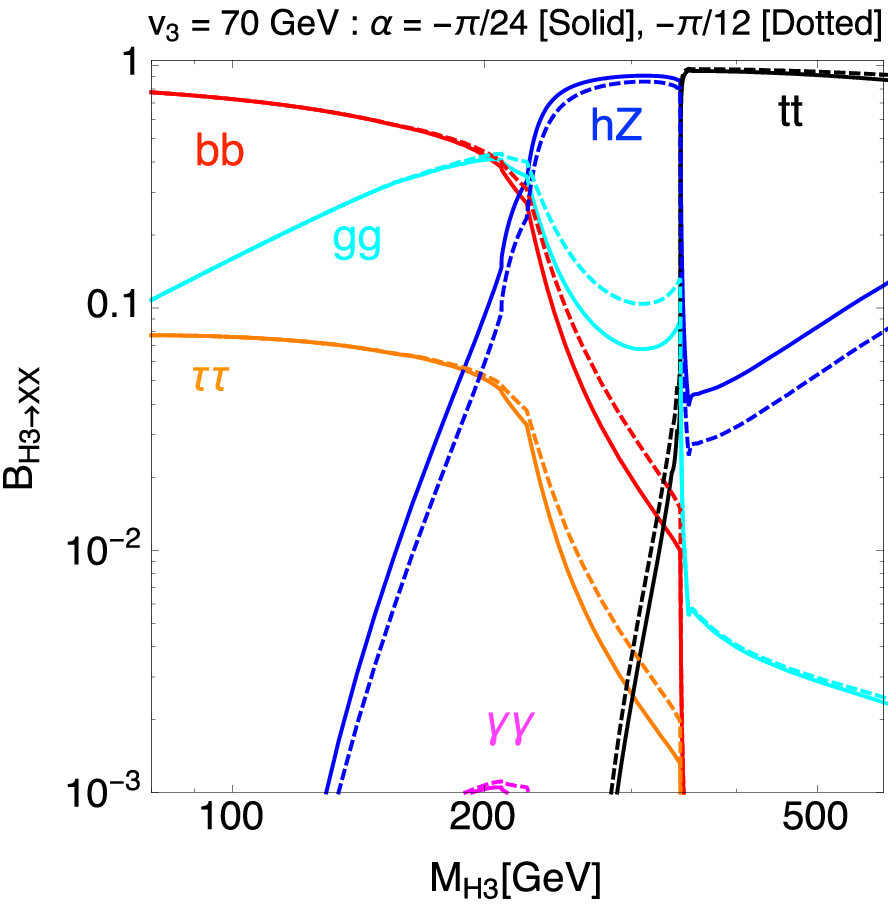}
\\
\vspace{-0.3cm}
(a) \hspace{4.75cm} (b) \hspace{4.75cm} (c)
\caption{Branching ratios of the $H_3^0$ boson as functions of $M_{H_3}$.  $v_3 = 10$ GeV, 30 GeV and 70 GeV in plots (a), (b), and (c), respectively.}
\label{FIG:BRH3}
\end{figure}

Note that the scaling factor for the couplings of $H_3^0$ to SM fermions is universally $\pm \cot\beta$, where the sign is determined by the sign of the third isospin number of the fermion.  Being a CP-odd particle, its fermion pair decay width is enhanced from that of a CP-even Higgs boson by the factor of $\left( 1 - 4M_f^2/M_{H_3}^2 \right)^{-1}$.  Besides, the $H_3^0$ boson does not couple to the weak bosons at tree level and therefore has no weak boson pair decay modes.  Moreover, its $gg$ and $\gamma\gamma$ decays are mediated only by fermions in the loop.  The particle has, however, a tree-level coupling with the SM-like Higgs boson and the $Z$ boson.  In fact, the $hZ$ mode becomes dominant once the decay is on mass shell, until the $t\bar t$ decay is also on shell.  The $H_3^0$-$h$-$Z$ coupling has dependences on both $\alpha$ and $\beta$ (or $v_3$).  The patterns of branching ratios as functions of the $H_3^0$ mass $M_{H_3}$ are shown in Fig.~\ref{FIG:BRH3}.

As shown in Fig.~\ref{FIG:BRH3}, the $b\bar b$ mode remains most dominant in the low-mass region up to the $hZ$ decay threshold.  When the $t\bar t$ decay is kinematically allowed, it becomes the most dominant mode.  The $hZ$ decay mode is important once again in the further higher mass region.  Incidentally, the branching ratio curve of the $\gamma\gamma$ mode has exactly the same behavior as the $gg$ curve, but down by the factor of $\left( Q_t \alpha_{\rm EM} / \alpha_S \right)^{2} \simeq 0.002$, where $\alpha_S$ and $\alpha_{\rm EM}$ denote respectively the strong and electromagnetic couplings and $Q_t = +2/3$.

Since the $H_3^0$ does not couple to the weak gauge bosons, one can only make use of the $f\bar f$ 
modes through the GGF production mechanism.  
In such cases, the signal strengths for the on-shell decays in the Born approximation are
\begin{align}
\mu_{FF}^\text{GGF} [H_3]
&=
(\kappa_F^{H_3})^2 
\frac{F_{1/2}^A(M_{H_3})}{F_{1/2}^S(M_{H_3})}
\times
\frac{{\mathcal B}_X}{{\mathcal B}_X^\text{SM}(M_{H_3})}
\left( 1 - \frac{4M_f^2}{M_{H_3}^2} \right)^{-1} ~,  
\end{align}
where $F_{1/2}^S(M)$ and $F_{1/2}^A(M)$ are the fermionic loop functions for CP-even and -odd scalar bosons~\cite{Ref:HHG}, respectively, and $(\kappa_F^{H_3})^2 = \cot^2\beta$, 
which goes to zero in the small $v_3$ limit.  Even though the above formula is only approximate, some general properties can be extracted.  Since the decays of a SM Higgs boson in the $M_{H_3} \alt 2 M_W$ region are dominated by the fermion pairs, $\mu_X^\text{GGF} [H_3] \sim \cot^2\beta$.  On the other hand, when $M_{H_3} > 2 M_W$, the inclusive ${{\mathcal B}_F^\text{SM}(M_{H_3})}$ is very small.  Therefore, the signal strengths of the fermionic modes have a significant increase at around $2M_W$.  This enhancement is slightly reduced when the $hZ$ mode opens up and further reduced above the $t\bar t$ threshold.

\begin{figure}[th]
\centering 
\includegraphics[width=7cm]{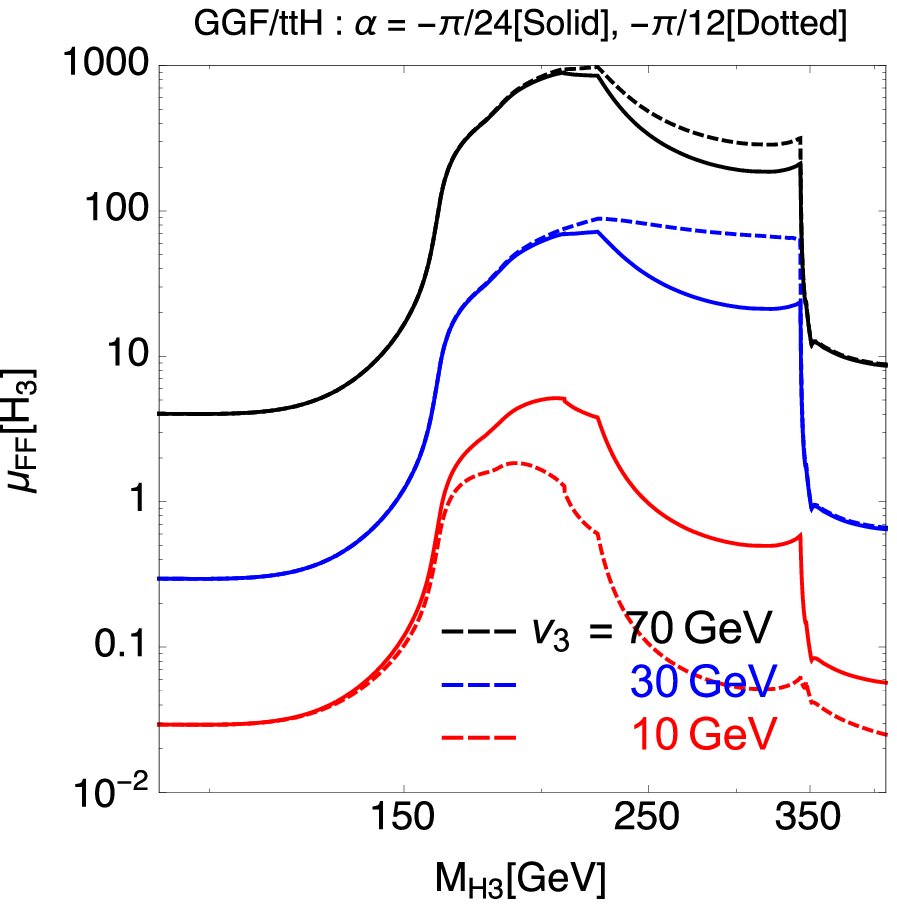}
\hspace{0.5cm}
\includegraphics[width=7cm]{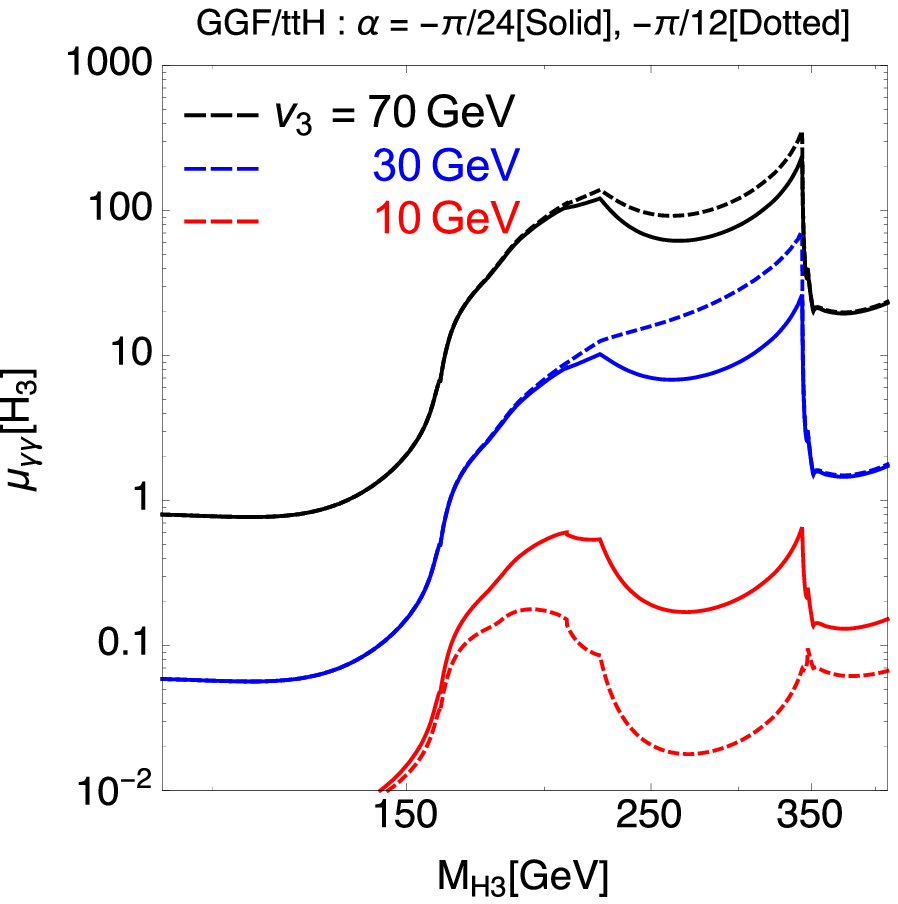}
\\
\vspace{-0.3cm}
(a) \hspace{7cm} (b)
\caption{Signal strengths of the $H_3^0$ boson in the GGF production of (a) the $f\bar f$ channels and (b) the $\gamma\gamma$ channel as functions of $M_{H_3}$.}
\label{FIG:SignalStrengthH3}
\end{figure}

In Fig.~\ref{FIG:SignalStrengthH3}, we show the signal strengths of the $f\bar f$ and $\gamma\gamma$ channels through the GGF production.  It is observed that the curves scale with $\cot^2\beta$.  Also, both channels have a significant enhancement in the mass range $2 M_W \alt M_{H_3} \alt M_Z + M_h$.  Therefore, a search of these GGF channels in this mass range can put stringent constraints on the model.  
Regarding the regions of $M_{H_3} \alt 2 M_W$ and $M_{H_3} \agt 2 M_t$,
the $f\bar f$ channel is smaller than the SM normalization when $v_3 \alt 30$ GeV.  Although having a signal strength greater than 1 for $v_3 \agt 30$ GeV and $M_{H_3} \agt 200$ GeV, the $\gamma\gamma$ channel is not useful as its production cross section is actually very small.

\subsection{The $H_5^0$ boson \label{sec:H5}}

\begin{figure}[th]
\centering 
\includegraphics[width=8cm]{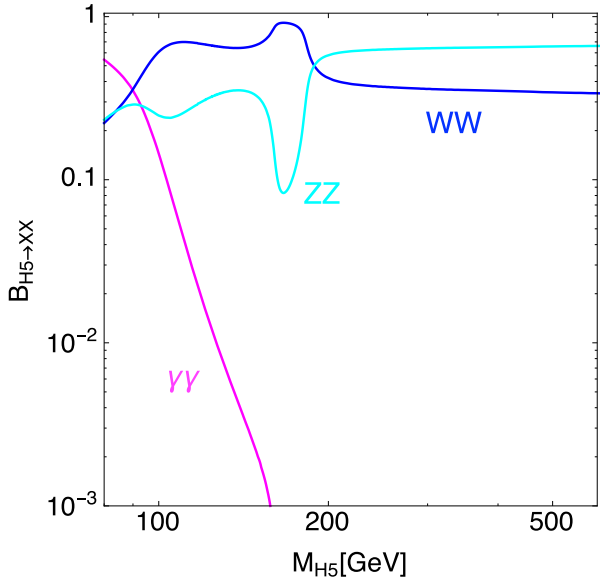}
\caption{Branching ratios of the $H_5^0$ boson as functions of $M_{H_5}$.}
\label{FIG:BRH5}
\end{figure}

Since the $H_5^0$ boson does not couple to the SM fermions at tree level, the only search channels are the $VV$ and $\gamma\gamma$ modes via the VBF production mechanism.  Note that as given in Table~\ref{tab:couplings}, both its couplings with the $W$ and $Z$ bosons scale with $\cos\beta$ (or equivalently $v_3$) without any dependence on $\alpha$.  But the latter is larger than the former by a factor of 2.  The pattern of branching ratios is independent of both $\alpha$ and $\beta$ (or $v_3$) and plotted as functions of the $H_5^0$ mass $M_{H_5}$ in Fig.~\ref{FIG:BRH5}.

The crossing between the curves for the $WW$ and $ZZ$ channels at both low-mass end and when $M_{H_5} \agt 190$ GeV is a result of the non-universal scaling behaviors in the couplings with the weak bosons. 
The ratio in the high mass region is $WW:ZZ = 1:2$ ({\it cf.} $2:1$ in the SM). 
This is a nice discriminant for the neutral Higgs boson originated from the custodial quintet. 
Because of the absence of fermionic decay modes, the fractions of $VV$ and $\gamma\gamma$ decay channels are much larger than in the SM.  Furthermore, the $\gamma\gamma$ decay channel can be significant for $M_{H_5} \lesssim 2M_W$ due to the off-shell suppression of $VV$ decay modes.  Again, contributions from the charged Higgs bosons to the $\gamma\gamma$ decay is neglected because we do not specify the Higgs potential explicitly.

\begin{figure}[th]
\centering 
\includegraphics[width=7cm]{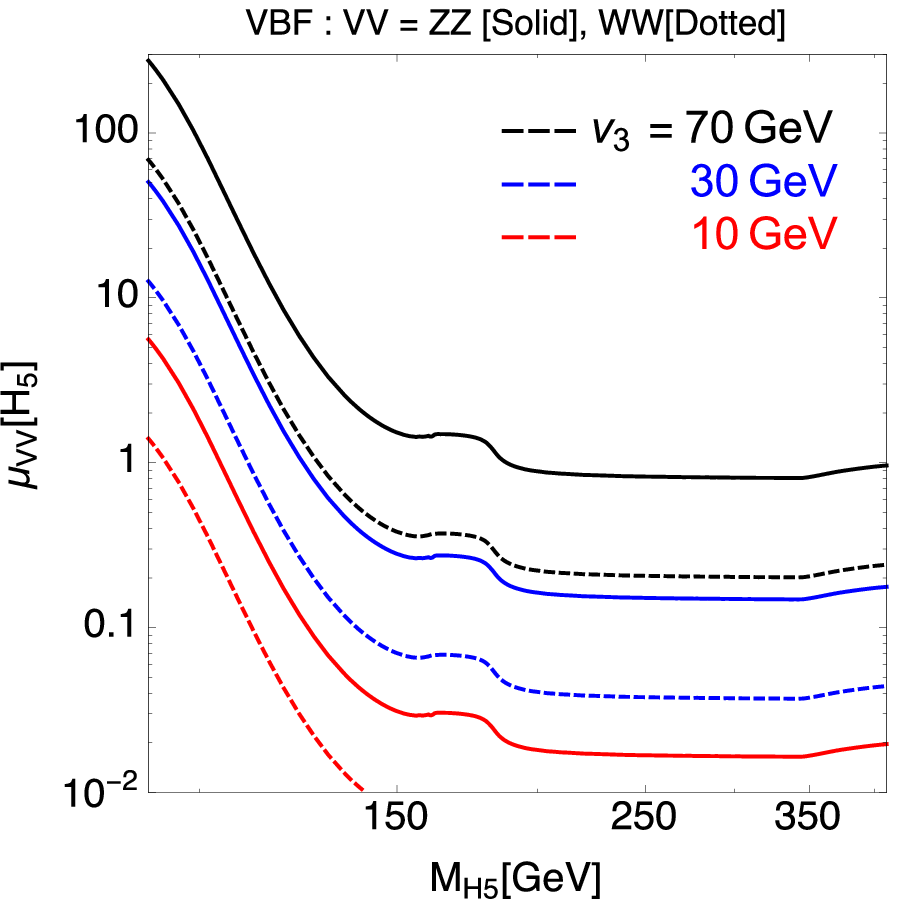}
\hspace{0.5cm}
\includegraphics[width=7cm]{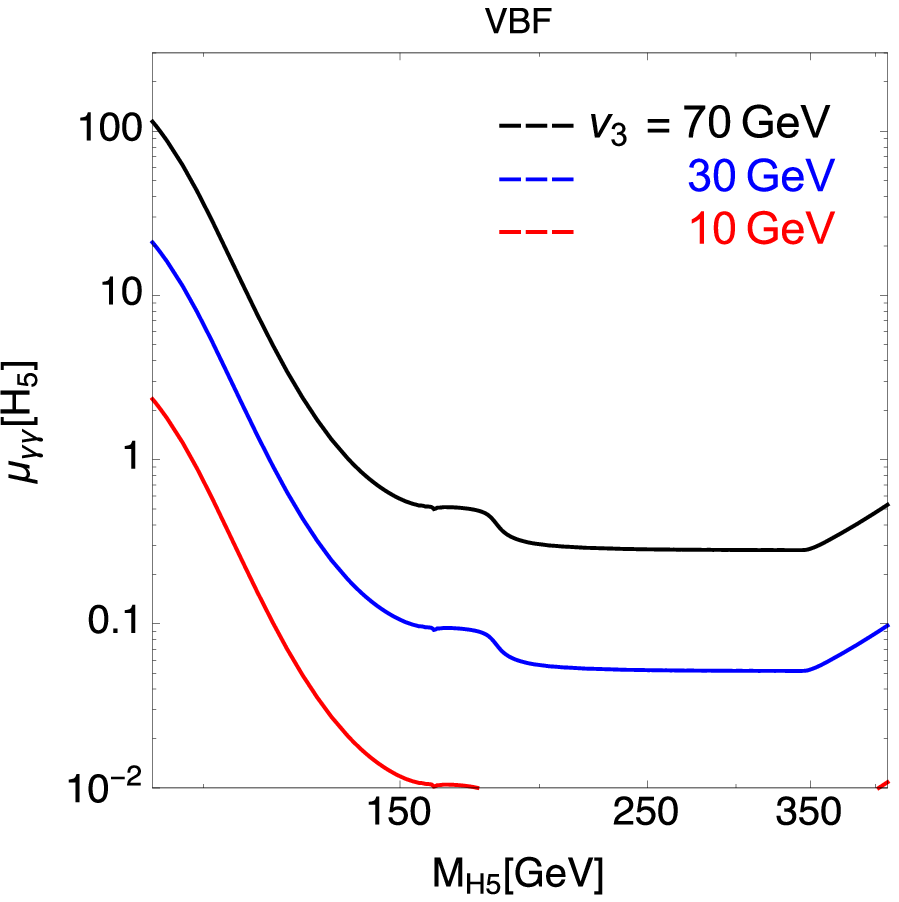}
\\
\vspace{-0.3cm}
(a) \hspace{7cm} (b)
\caption{Signal strengths of the $H_5^0$ boson as functions of $M_{H_5}$.
}
\label{FIG:SignalStrengthH5}
\end{figure}

The signal strengths of the $ZZ$, $WW$, and $\gamma\gamma$ modes through the VBF mechanism are plotted in Fig.~\ref{FIG:SignalStrengthH5}.  
Note that there is no $\alpha$ dependence in the curves, and the $v_3$ dependence comes from the production part as the decay branching ratios are independent of both $\alpha$ and $\beta$. 
The VBF cross section given by the Higgs Cross Section Working Group~\cite{Ref:XSWG} is the sum of $WW$ fusion and $ZZ$ fusion cross sections.  In order to reflect their respective modifications in the production part, we have evaluated separately the $WW$ fusion and $ZZ$ fusion cross sections using MadGraph5\_aMC@NLO~\cite{Ref:MG5}.  The ratio is found to be about $3:1$ for a wide range of $M_{H_5}$.
As expected from the prediction of branching ratios, the signal strengths for $VV$ and $\gamma\gamma$ decay channels are both enhanced substantially in the low mass region, $M_{H_5} \lesssim 2M_W$.

\section{Experimental Constraints \label{sec:constraints}}

Searches of another SM-like Higgs boson with a different mass have been done at the LHC for various channels.  Upper limits at 95\% confidence level (CL) are provided as a function of the extra Higgs mass in a wide range, from as small as $60$ GeV up to 1 TeV depending on the search channels.  We can then take such data as constraints on the searches of the exotic neutral Higgs bosons in the GM model.

In analyzing the experimental data, we find that the ATLAS results are particularly convenient to use because they are all taken from the 8-TeV run with integrated luminosities about 20 fb$^{-1}$, whereas the CMS results mix measurements in both 7-TeV and 8-TeV runs.  Therefore, we will use exclusively the ATLAS results in the following analysis.  Up to now, the $H \to ZZ \to 4\ell$ mode is analyzed separately in both GGF and VBF+VH production processes~\cite{Ref:ZZ4l}, and the $H \to WW \to e\nu \mu\nu$ mode in both GGF and VBF production processes~\cite{Ref:WWevmv}.  The $H \to \gamma\gamma$ mode is divided into low-mass ($65$--$110$ GeV) and high-mass ($110$--$600$ GeV) regions~\cite{Ref:GamGam}.  Since no similar data analysis using the fermionic modes is available yet, currently we can only extract constraints on the model through the studies of the $H_1^0$ and $H_5^0$ bosons.  However, as we noted in Section~\ref{sec:H3}, the fermionic channels in the mass region of $2M_W \alt M_{H_3} \alt M_Z + M_h$ can easily put stringent constraints on the model parameter space.

\begin{figure}[th]
\centering 
\includegraphics[width=7cm]{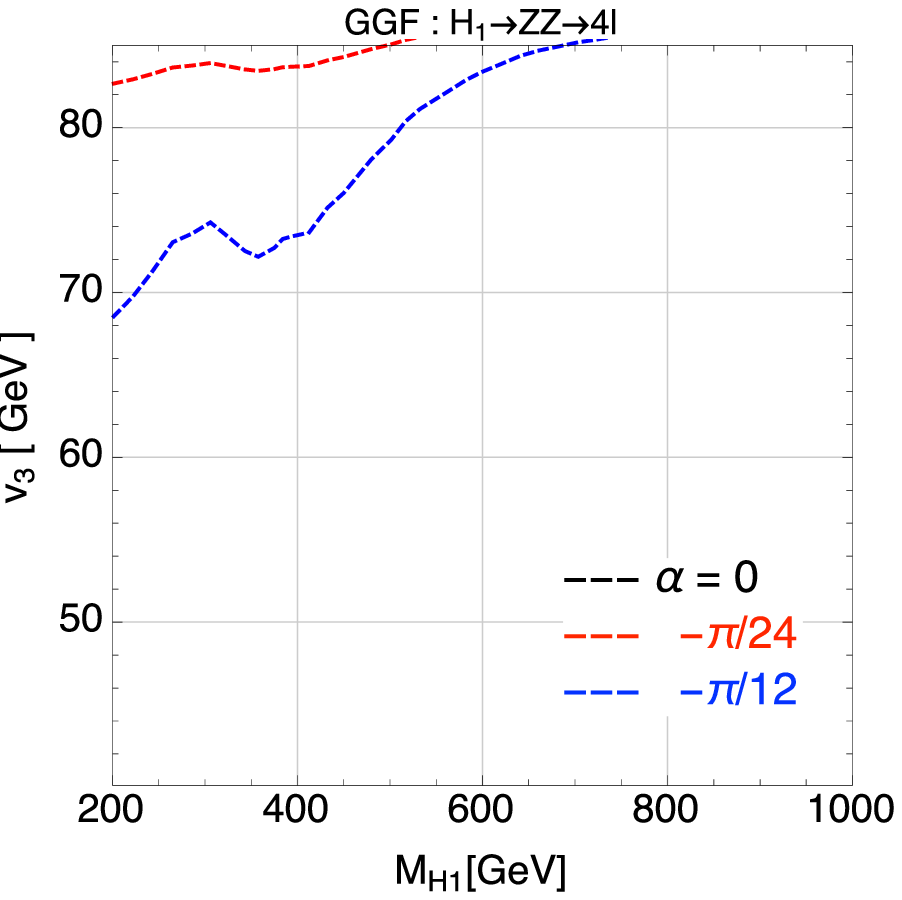}
\hspace{0.5cm}
\includegraphics[width=7cm]{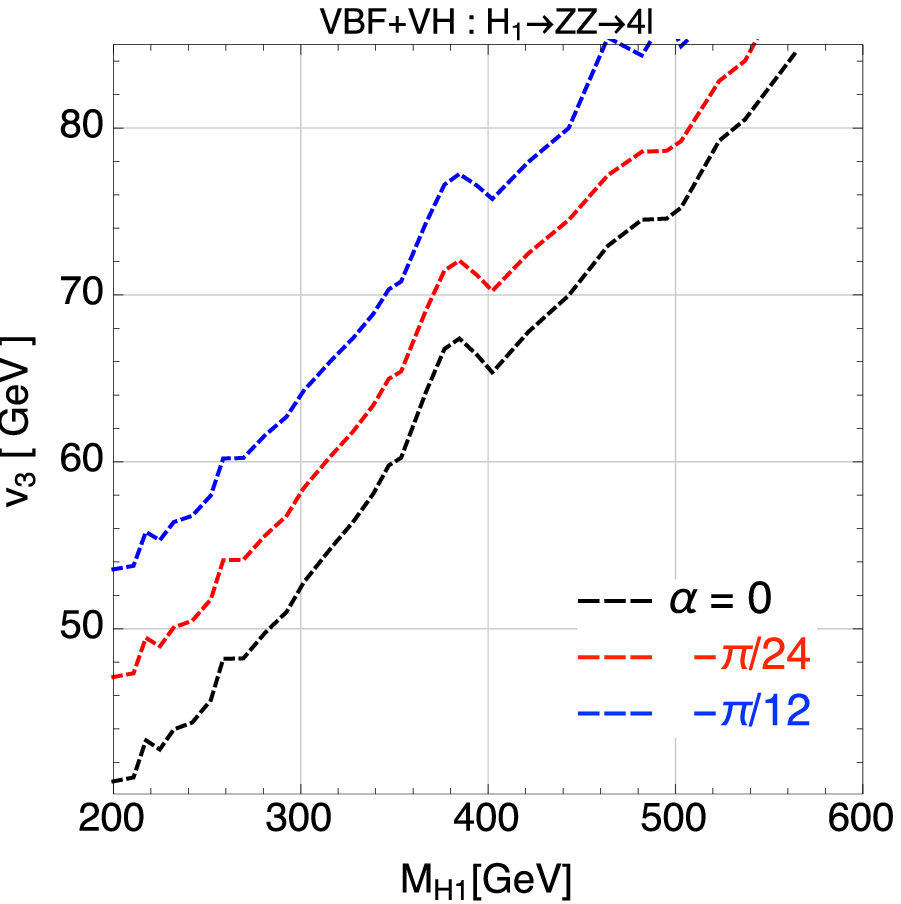}
\\
\vspace{-0.2cm}
(a) \hspace{7cm} (b)
\\
\bigskip
\includegraphics[width=7cm]{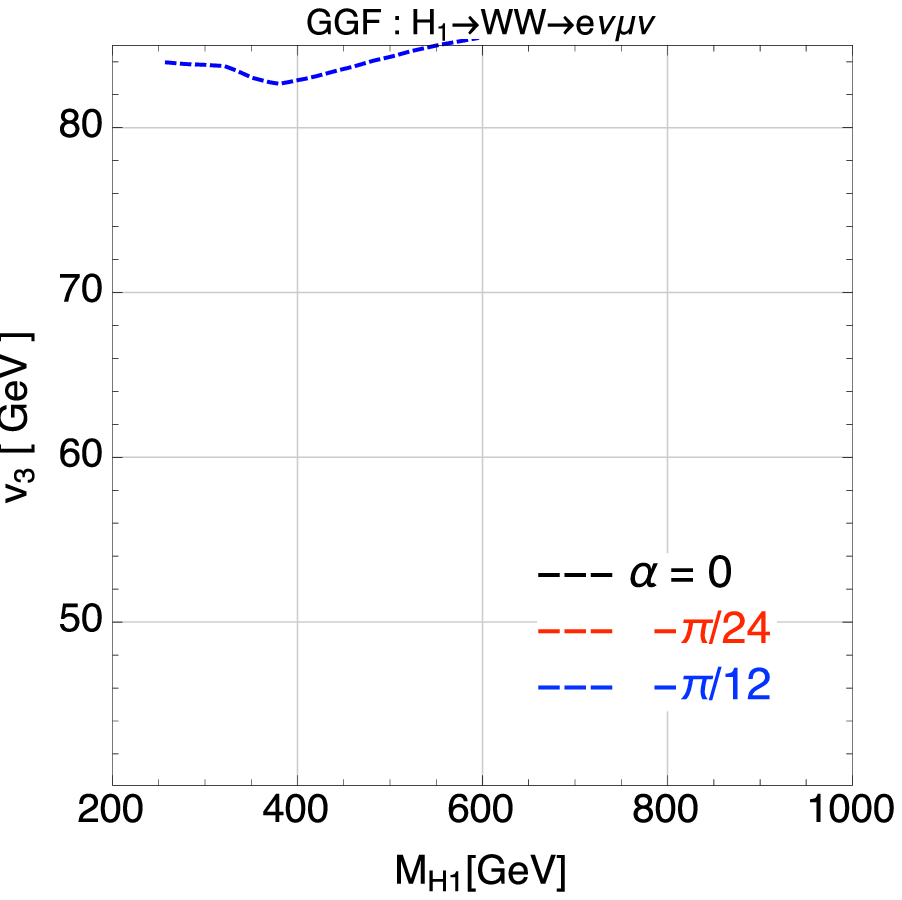}
\hspace{0.5cm}
\includegraphics[width=7cm]{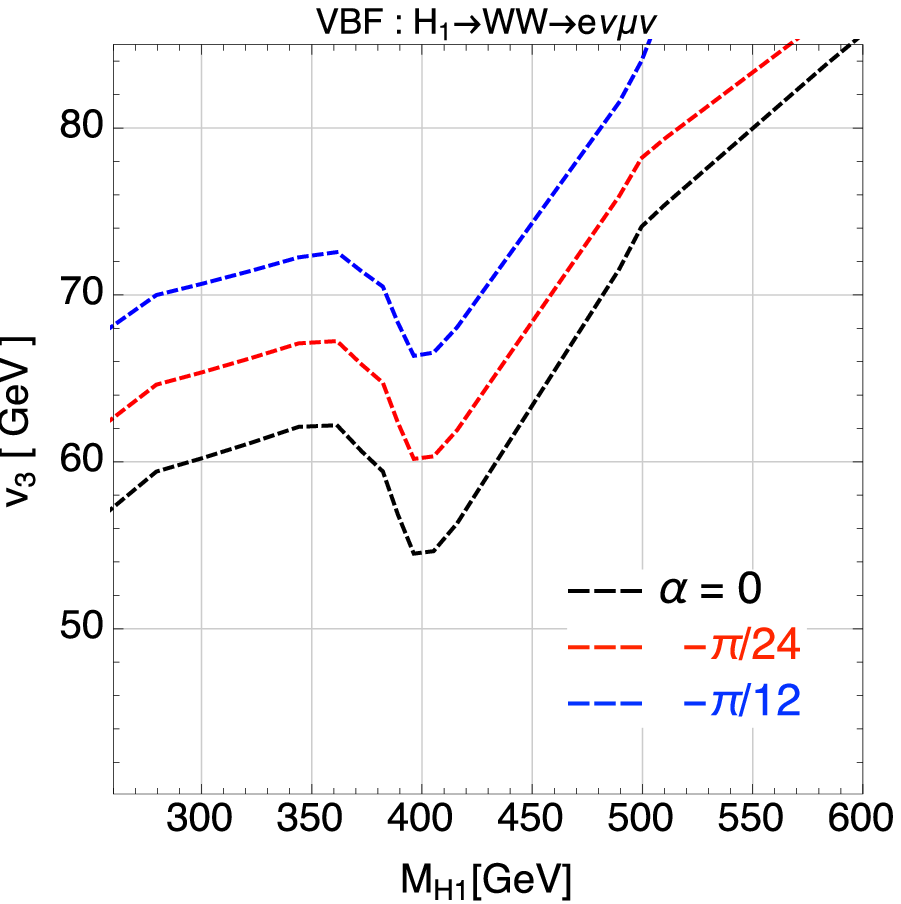}
\\
\vspace{-0.2cm}
(c) \hspace{7cm} (d)
\caption{Upper limits on $v_3$ as a function of $M_{H_1}$ for different channels and production mechanisms.  Plot (a) involves the GGF mechanism, plot (b) the VBF+VH mechanism, plot (c) the GGF mechanism, and plot (d) the VBF mechanism.  From top to bottom are for the $WW$ and $ZZ$ modes.}
\label{FIG:Limit_on_v3_MH}
\end{figure}

Upper limits on the triplet VEV at 95\% CL from the $ZZ$ and $WW$ modes are plotted against the $H_1^0$ mass in Fig.~\ref{FIG:Limit_on_v3_MH}.  It is clear that in both channels, the VBF process has a stronger constraint than the GGF process.  This is consistent with the earlier observation that the $H_1^0$ boson is fermiophobic in the small $\alpha$ scenario considered here.  Comparing Fig.~\ref{FIG:Limit_on_v3_MH}(b) and Fig.~\ref{FIG:Limit_on_v3_MH}(d), one notices that the $ZZ$ channel is generally more constraining than the $WW$ channel except for the region $375 \alt M_{H_1} \alt 450$ GeV, in which the former (latter) has a slightly worse (better) sensitivity experimentally.

\begin{figure}[th]
\centering 
\includegraphics[width=8cm]{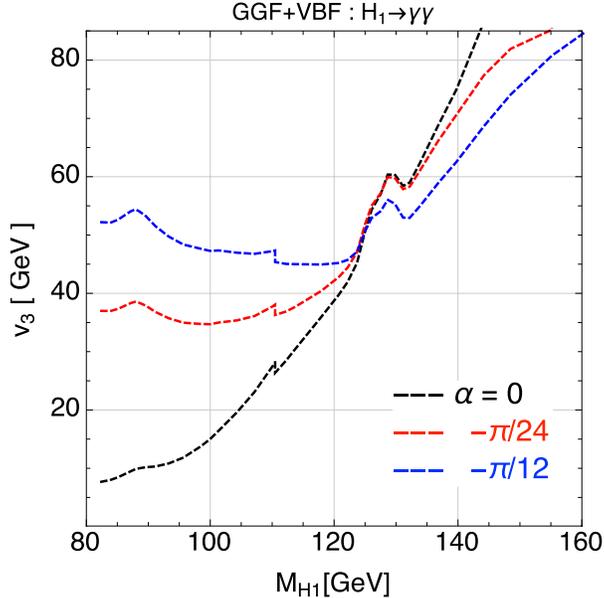}
\caption{Upper limits on $v_3$ as a function of $M_{H_1}$ for the $\gamma\gamma$ channels using events produced through the GGF, VBF, and VH mechanisms.
}
\label{FIG:Limit_on_v3_MH_GamGam}
\end{figure}

In Fig.~\ref{FIG:Limit_on_v3_MH_GamGam}, we plot the upper limit of the triplet VEV as a function of $M_{H_1}$ from the $\gamma\gamma$ decay mode. 
In order to compare with the experimental data, we need to have an efficiency factor $C_X$~\cite{Ref:GamGam}, which is contained in the calculation of the fiducial cross section.  The mass dependence of $C_X$ is not presented in Ref.~\cite{Ref:GamGam}; only the range of $0.56$ to $0.71$ for the $C_X$ factor is mentioned.  We here take $C_X=0.56$ to calculate a {\it conservative} limit on the triplet VEV. 
As shown in Fig.~\ref{FIG:Limit_on_v3_MH_GamGam}, the data are more (less) constraining for smaller $\alpha$ in the lower (higher) mass regime.

\begin{figure}[th]
\centering 
\includegraphics[width=8cm]{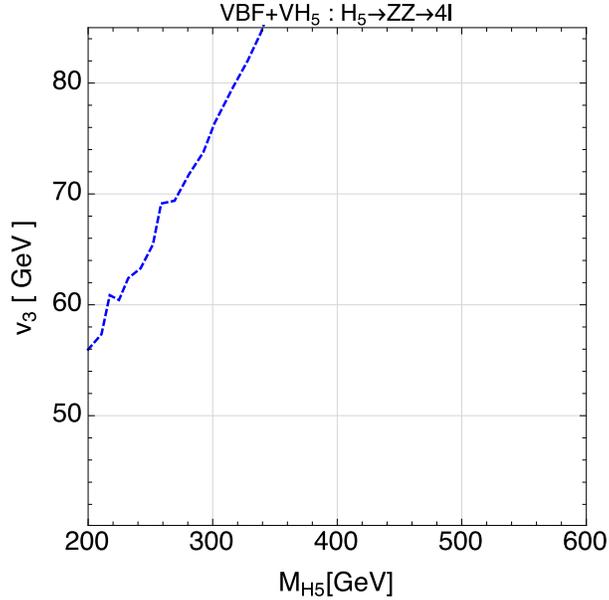}
\caption{Upper limits on $v_3$ as a function of $M_{H_5}$ for the $ZZ \to 4 \ell$ channels through the VBF mechanism.
}
\label{FIG:Limit_on_v3_MH5}
\end{figure}

\begin{figure}[th]
\centering 
\includegraphics[width=8cm]{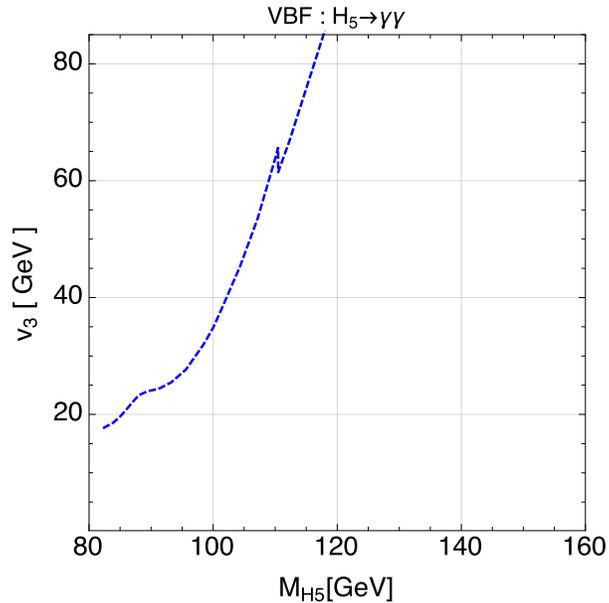}
\caption{Upper limits on $v_3$ as a function of $M_{H_5}$ for the $\gamma\gamma$ channels through the VBF mechanism.
}
\label{FIG:Limit_on_v3_MH5_GamGam}
\end{figure}

Finally, Fig.~\ref{FIG:Limit_on_v3_MH5} shows the upper limit of the triplet VEV as a function of $M_{H_5}$ from the $ZZ \to 4 \ell$ decay mode via the VBF mechanism.  Clearly, this constraint from the search of $H_5^0$ is weaker than those presented in Figs.~\ref{FIG:Limit_on_v3_MH} and \ref{FIG:Limit_on_v3_MH_GamGam}, where we focus on the small mixing scenario ($-\pi/12 \le \alpha \le 0$).  This is related to the fact that the signal strength is mainly enhanced in the low-mass region only.  We note in passing that no useful constraint can be obtained from the $WW$ mode. 
Similar to $H_1^0 \to \gamma\gamma$, one can also extract constraints from the $H_5^0\to \gamma\gamma$ mode, as shown in Fig.~\ref{FIG:Limit_on_v3_MH5_GamGam}. 
The $H_5^0\to \gamma\gamma$ channel may pose a stringent bound on $v_3$ in the small $M_{H_5}$ region.  However, there is an uncertainty in the Higgs trilinear couplings omitted in this analysis.

\section{Summary \label{sec:summary}}

The Georgi-Machacek model predicts the existence of four neutral Higgs bosons, one of which can be identified as the discovered SM-like Higgs boson $h$ and the other three $H_1^0$, $H_3^0$ and $H_5^0$ belong respectively to custodial singlet, triplet and quintet.  In addition to restraining the model parameters using the SM-like Higgs data, one can use the data of searching for additional neutral Higgs bosons at other masses to put constraints.  Taking the cue from current SM-like Higgs data, we focus on the parameter space of small mixing angle $\alpha$ and study the decay branching ratios and signal strengths of these Higgs bosons at the CERN LHC.  We have examined in this work how they vary with $\alpha$, the triplet VEV $v_3$, and the corresponding Higgs mass.

As a result of mixing with $h$, $H_1^0$ share all the decay modes of $h$.  Moreover, in the small $\alpha$ scenario considered in this work, the weak boson decay modes are dominant in the high-mass region for large triplet VEV $v_3$.  Due to its odd CP property, the $H_3^0$ does not have the weak boson channels.  Its dominant decay mode from the low-mass region up to about 600 GeV is $b \bar b$, $hZ$, and $t \bar t$.  In the case of $H_5^0$, it mainly decays to the $ZZ$, $WW$, and $\gamma\gamma$ states.  In the high-mass region, the branching ratio of $ZZ$ is about twice that of $WW$.

Using the latest search data of extra Higgs boson from ATLAS, we have put constraints on $v_3$ as a function of the Higgs mass for the $H_1^0$ and $H_5^0$ bosons.  For $H_1^0$, we have employed the $ZZ \to 4 \ell$, $WW \to e\nu \mu\nu$, and $\gamma\gamma$ modes through the gluon-gluon fusion (GGF) and vector boson fusion (VBF) productions.  The constraints from the VBF mechanism are stronger than those from the GGF mechanism, and the results have dependence on the $\alpha$ angle.  For $H_5^0$, we have found that only the $ZZ \to 4 \ell$ mode is useful and renders a weaker constraint than those of $H_1^0$.  Nevertheless, this constraint is independent of $\alpha$.

We have found that the fermionic channels of $H_3^0$ are enhanced a lot in the mass range between $2 M_W$ and $2M_t$.  Therefore, a search of such channels in this regime can readily discover the particle or put stringent constraints on the model parameters.

\section*{Acknowledgments}
This research was supported in part by the Ministry of Science and Technology of Taiwan under Grant No.\ NSC 100-2628-M-008-003-MY4 and in part by the MEXT Grant-in-Aid for Scientific Research on Innovative Areas No.\ 26104704.  C.W.C. would like to thank the hospitality of the Kobayashi-Maskawa Institute at Nagoya University where this work was done during his sabbatical visit.  He also thanks Chia-Ming Kuo for detailed explanations of some experimental data.


\end{document}